\documentclass[epj]{svjour}
\usepackage{graphics}
\usepackage{caption}
\usepackage{subcaption}
\RequirePackage[T1]{fontenc}
\usepackage{float}
\RequirePackage{graphicx}
\RequirePackage{mathptmx}      
\RequirePackage{flushend}
\RequirePackage[numbers,sort&compress]{natbib}
\RequirePackage[colorlinks,citecolor=blue,urlcolor=blue,linkcolor=blue]{hyperref}

\usepackage[numbers]{natbib}
\usepackage{rotating}

\usepackage{graphicx}
\usepackage{braket}
\usepackage{multirow}
\usepackage{amsmath}
\begin{document}
\title{Investigation of Mass and Decay Characteristics of the All-light Tetraquark }
\author{Chetan Lodha\inst{1},
	 \and  Ajay Kumar Rai\inst{2},}                     

%
\institute{Department of Physics, Sardar Vallabhbhai National Institute of Technology, Surat, Gujarat-395007, India, \\ \email{iamchetanlodha@gmail.com }\\ \email{raiajayk@gmail.com}}

\mail{iamchetanlodha@gmail.com}
\date{Received: date / Revised version: date}
%
\abstract{	We investigate the mass spectra and decay properties of pions and all light tetraquarks using both semi-relativistic and non-relativistic frameworks. By applying a Cornell-like potential and a spin-dependent potential, we generate the mass spectra. The decay properties of tetraquarks are evaluated using the annihilation model and the spectator model. Potential tetraquark candidates are interpreted for quantum numbers $J^{PC} = 0^{++}, 0^{-+}, 1^{-+}, 1^{+-}, 1^{--}, 2^{+-}, 2^{-+},$ and $2^{--}$. Additionally, we compare our results with existing experimental data and theoretical predictions to validate our findings. This study aims to enhance the understanding of tetraquarks in the light-light sector.}

\maketitle
\section{Introduction}
Exotic hadrons represent a fascinating and complex aspect of strong interactions as described by quantum chromodynamics (QCD) \cite{Brambilla:2019esw}. While conventional hadrons (mesons and baryons) have been well-categorized and studied in detail\cite{ Chaturvedi:2022pmn,Purohit:2022mwu,Menapara:2023rur}, exotic hadrons open a new window to understanding the multi-body dynamics of gluons and quarks \cite{Gell-Mann:1964ewy}. Among these, tetraquarks are one of the most well-known types and were first proposed theoretically in the 1970s \cite{Jaffe:1976ig,Jaffe:1976yi}. The first experimental evidence for a potential tetraquark emerged in 2003, marking the beginning of a new era in the study of exotic hadrons \cite{Belle:2003nnu}. Since then, numerous tetraquark candidates have been observed at facilities such as Belle, LHCb, and  $D\emptyset$. 
Various pieces of literature have investigated the possible structures of four-quark resonances, interpreting them as either compact tetraquarks or mesonic molecules, also known as molecular tetraquarks, in singlet states \cite{Brambilla:2019esw,Chen:2022asf}. Several notable tetraquark candidates have been identified in experimental studies, including Z(4430) by Belle \cite{Belle:2007hrb}, Y(4660) by Belle, Y(4140) by Fermilab \cite{Mahajan:2009pj}, X(5568) by the $D\emptyset$ experiment \cite{D0:2016mwd}, $Z_{c}(3900)$ by BESIII \cite{BESIII:2013ris}, and X(6900) by LHCb \cite{LHCb:2020bwg}. While most experimentally discovered tetraquark candidates contain at least one heavy quark, numerous theoretical studies over the past decade have proposed various resonances for all-light tetraquarks. Notably, $f_{0}(980)$, $a_{0}(980)$, and $f_{0}(500)$ (formerly known as $\sigma$) have emerged as strong candidates for all-light tetraquarks.
The $\pi$ meson, or pion, was the first meson to be both predicted and detected, with its experimental discovery in 1947 marking a significant milestone in particle physics \cite{Lattes:1947mw}. Since then, extensive research has been conducted on pions, greatly enhancing our understanding of mesons and other hadrons. Over the past decade, pion physics has seen remarkable progress, with numerous studies and experiments shedding light on its nature and dynamics. Significant strides have been made by experiments such as COMPASS and BESIII, focusing on detecting ultra-rare pion decay processes \cite{Xian:2022icn,Yao:2024pij}. In heavy-ion collision experiments at the LHC and RHIC, pions are the most abundantly produced particles. These collisions create extreme conditions similar to those shortly after the Big Bang, forming a quark-gluon plasma \cite{LHCb:2012myk}. Measurements of pion spectra and their correlations in these experiments have provided new insights into the properties of this exotic state of matter. Recently, numerous resonances have been observed in the 1-2.5 GeV mass range, which resembles the nature of pions \cite{Chen:2022qpd,E862:2006cfp,BES:2005ega,BESIII:2021xoh}. These efforts have deepened our understanding of the strong nuclear force, the internal structure of pions, and the properties of the early universe. These experimental updates will be helpful in understanding the nature of the unconfirmed strange meson listed in PDG. In order to explain the structure of hadrons, multiple theoretical models have been developed over the years. \cite{Rai:2008sc,Kher:2017wsq,Kher:2017mky,Kher:2018wtv,Oudichhya:2023lva}. The current study calculates the mass spectra and $J^{PC}$ values for the S, P, D, F, and G waves of pions. 
Studies describing compact tetraquarks often characterize them as bound states of diquarks and anti-diquarks. The present work determines the meson fitting parameters and then uses them to calculate the properties of diquarks and tetraquarks. The heavy tetraquark system has been explored using various models such as the Bag model, lattice QCD \cite{Wagner:2013vaa}, QCD Sum Rule \cite{Chen:2016jxd,Agaev:2024pej}, and potential phenomenology models \cite{Ghalenovi:2020zen,Noh:2023zoq,Lodha:2023gpp}. Similarly, the mass spectra of other exotic states, including hadronic molecules, pentaquarks, and hybrid mesons, have been studied in detail using non-relativistic formalisms \cite{Rai:2006bt}.
Numerous resonances in the mass range of 1.0–3.5 GeV have been observed at experimental facilities, displaying characteristics indicative of four-quark states. Several studies suggest that $\eta'$, $K^{*}(700)$, $f_{0}(980)$, $a_{0}(980)$, $f_{0}(1370)$, and $f_{0}(1500)$ are better interpreted as four-quark states rather than traditional two-quark states. For instance, Ref \cite{Kim:2023bac,Kim:2024adb} proposes a tetraquark mixing model that treats both light and heavy nonets as tetraquarks formed by the mixing of two tetraquarks. An extensive study on the $qq\bar{q}\bar{q}$ structure is provided in study \cite{Santopinto:2006my}, which predicts a tetraquark nonet including $f_{0}(600)$, $\kappa(800)$, $f_{0}(980)$, and $a_{0}(980)$. Using the QCD Sum Rule, Ref. \cite{Wang:2006gj} investigates X(1576) as a tetraquark state in the P-wave with scalar diquarks $[us]$, $[ds]$, $[\bar{u}\bar{s}]$, and $[\bar{d}\bar{s}]$. Similarly, study \cite{Wang:2019nln} employs the QCD Sum Rule to conduct a comprehensive study on fully light vector tetraquark states explicitly in S-wave tetraquarks and extends this study to P-wave tetraquarks in Ref \cite{Xin:2022qnv}. Study \cite{Zhao:2021jss} calculates the masses of the $qq\bar{q}\bar{q}$ tetraquark ground state and first radial excited state using a constituent quark model that incorporates a Cornell-like potential and one-gluon exchange spin-spin coupling. Ref. \cite{Wang:2024pgy} explores light tetraquark states with exotic quantum numbers $J^{PC} = 2^{+-}$ using the QCD Sum Rule. Ref. \cite{Vijande:2009ac} performs a detailed analysis of the symmetry properties of a four-quark wave function, solving it via a variational approach to investigate several resonances as potential all-light tetraquark candidates. Masses of the ground-state light tetraquarks are dynamically calculated in Ref. \cite{Ebert:2008id} using the relativistic diquark-antidiquark framework. Using the QCD two-point sum rule approach, Ref. \cite{Agaev:2018fvz} calculates the spectroscopic parameters and partial decay widths of the light meson $a_{0}(980)$ by treating it as a scalar diquark-antidiquark state. The present study calculates the mass spectra of S-wave and P-wave $qq\bar{q}\bar{q}$ tetraquarks and explores their decay properties along with possible decay channels.

\begin{figure}
	\centering
	\includegraphics[width=0.5\linewidth, height=0.25\textheight]{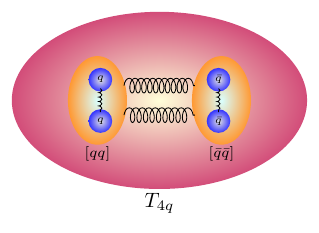}
	\caption{Pictorial representation of all light tetraquark}
	\label{fig:tikz}
\end{figure}

The present paper is organized as follows: Section II presents the theoretical framework for determining the mass spectra of mesons, diquarks, and tetraquarks, after a brief introduction in Section I. The mass spectroscopy of diquarks, anti-diquarks, tetraquarks, and pion  is explained and examined in Section III. The decay characteristics of tetraquarks are addressed in Section IV. The Regge trajectories of pions and tetraquarks are examined in Section V. We present the findings and discussion in Section VI, with a conclusion in Section VII.

\section{Framework}
Building upon the theoretical description of quark interactions and their potential phenomenology, the present research adopts the methodology outlined by \cite{Lodha:2024qby}. This study employs a combination of semi-relativistic and non-relativistic models to characterize the quark composition and internal structure of tetraquarks. {The mass of the $\rho$ meson is approximately 0.77 GeV, while the diquark considered in this study has a mass close to 1 GeV, which serves as a threshold between relativistic and non-relativistic regimes. Composite particles with masses below 1 GeV are more accurately described using relativistic mechanisms, whereas those in the 1-2.5 GeV range can be effectively modeled with either semi-relativistic or non-relativistic formalisms. Although the non-relativistic framework has been widely applied to heavy systems, such as all-heavy tetraquarks and heavy quarkonia, its application to light diquarks remains relatively underexplored. This gap presents an opportunity to extend the relativistic formalism to better understand the dynamics and structure of light diquarks, potentially leading to new insights into their behavior.} Central to this endeavor is the computation of binding energies associated with distinct states, accomplished through the utilization of a modified time-independent radial Schrödinger equation, with detailed discussions provided in reference \cite{Tiwari:2021iqu,Tiwari:2021tmz}. The integration of the two-body problem into a central body problem is achieved utilizing the center of mass frame, with a comprehensive exposition available in \cite{Lucha:1995zv}. To delve deeper into the methodology, the study isolates the radial and angular components of the time-independent Schrödinger equation by employing spherical harmonics. Specifically pertaining to mesons and tetraquarks, the foundational two-body semi-relativistic and non-relativistic Hamiltonian in the center of mass frame accommodates the relativistic motion of constituent particles within the bound state, as outlined in \cite{Devlani:2011zz,Esposito:2016noz} and is given by,

\begin{equation}
	H_{SR} = \sum_{i=1}^2 \sqrt{M^{2}_{i} + p^{2}_{i}+} + V^{(0)} (r) + V_{SD}(r),
\end{equation}

\begin{equation}
	H_{NR} = \sum_{i=1}^2 (\frac{p_{i}^{2}}{2M_{i}} + M_{i}) + V^{(0)} (r) + V_{SD}(r).
\end{equation}

where $M_{i}$, $p_{i}$, $V^{0} (r)$ and $V_{SD}(r)$ are the constituent mass, relative momentum of the bound state system, interaction potential and spin-dependent potential respectively. The kinetic energy term of the Hamiltonian is expanded up to $\mathcal{O}(p^{10})$, signifying a rigorous treatment that accounts for higher-order momentum contributions. This expansion allows for a more precise description of the dynamics, capturing subtleties that might otherwise be overlooked in lower-order approximations.
 {In the expansion up to \( O(p^{10}) \), the terms \( O(p^4) \) and \( O(p^8) \) exhibit negative values, while \( O(p^6) \) and \( O(p^{10}) \) display positive values. This alternating polarity among the terms suggests a potential cancellation effect, where the positive and negative contributions may partially negate each other. This phenomenon highlights the intricate interplay within the expansion, indicating that certain terms might counterbalance the effects of others, leading to a more refined and accurate description of the system. Notably, the \( O(p^{10}) \) term plays a critical role in shaping the mass spectra of all-light tetraquarks. Its significance becomes evident when considering the delicate balance between the positive and negative contributions, which collectively influence the overall behavior of these particles. The presence of this higher-order term suggests that the expansion's convergence and the resulting mass spectra are particularly sensitive to the inclusion of higher-order contributions. The nuanced dynamics revealed through this balance underscore the importance of carefully considering all relevant terms in the expansion, in accurately capturing the mass distribution of all light tetraquarks.} This is particularly crucial when investigating systems involving high-energy quarks and complex quark interactions.  For further insights and discussions on the implications of this expanded kinetic energy term, refer to the comprehensive analysis presented in reference \cite{Patel:2022hhl}. This source elaborates on the theoretical underpinnings and practical implications of extending the kinetic energy expansion to $\mathcal{O}(p^{10})$, shedding light on its role in refining our understanding of tetraquark structures and their behavior within the quantum chromodynamics framework. The kinetic energy term is given as:
\begin{equation} 
	\begin{split}
		K.E. = & \sum_{i=1}^2 \frac{p^{2}}{2}\bigr(\frac{1}{M_{i}}\bigr) - \frac{p^{4}}{8}\bigr(\frac{1}{M_{i}^{3}}\bigr) + \frac{p^{6}}{16}\bigr(\frac{1}{M_{i}^{5}}\bigr) \\ 
		&- \frac{5p^{8}}{128}\bigr(\frac{1}{M_{i}^{7}}\bigr) + \frac{7p^{10}}{256}\bigr(\frac{1}{M_{i}^{9}}\bigr)
	\end{split}
\end{equation}

Numerous studies have extensively explored the hadron spectrum using different potentials, with the zeroth-order Cornell potential emerging as a fundamental model. This potential is characterized by a coulombic term resulting from one-gluon exchange, represented as a Lorentz vector exchange, and a linear term arising from quark confinement, represented by a Lorentz-Scalar exchange. The Coulombic term accounts for the short-range interactions mediated by gluons, reflecting the attractive force between quarks due to the exchange of color-charged particles. Meanwhile, the linear term embodies the long-range confinement force that increases linearly with the separation of quarks, illustrating the phenomenon of quark confinement within hadrons. This Cornell potential has proven instrumental in elucidating the energy spectra and properties of various hadronic systems, providing crucial insights into the underlying dynamics governed by quantum chromodynamics. The zeroth-order Cornell-like potential is given by 
\begin{equation}
	V^{(0)}_{C+L}(r) = \frac{k_{s}\alpha_{s}}{r} + br + V_{0},
\end{equation}

where $\alpha_{s}$, $k_{s}$, $b$ and $V_{0}$ are the QCD running coupling constant, color factor, string tension and constant, respectively.

Different color configurations of hadronic states correspond to distinct color factors that characterize their color charge interactions. Specifically, singlet states, representing a color-neutral configuration, are associated with a color factor, $k_{s}$  of $-\frac{4}{3}$. Triplet-antitriplet states, which involve a combination of a color triplet and its corresponding antitriplet, have a color factor, $k_{s}$ of $-\frac{2}{3}$. On the other hand, sextet states, which possess a more complex color structure, exhibit a color factor, $k_{s}$ of $\frac{1}{3}$. These color factors reflect the relative strength and nature of the color interactions within these states, providing essential information about their quantum properties and behavior in the framework of QCD. {The color factor and its physics are studied in detail in ref. \cite{Muta:2010xua,Deba}.} The incorporation of relativistic corrections to masses within the central potential framework draws inspiration from the work of \cite{Koma:2006si} and revises the central potential as,
\begin{equation}
	V^{(0)}(r) = V^{0}_{C+L}(r) + V^{1}(r) \biggl(\frac{1}{m_{1}} + \frac{1}{m_{2}} \biggl)
\end{equation}

where $m_{1}$ and $m_{2}$ denote the constituent masses of constituent particles in the bound state. Since the non-perturbative form of the relativistic mass correction term remains insufficiently explored, the leading order of perturbation theory is utilized. Specifically, this involves employing the first-order approximation within perturbative methods. The expression for this leading-order correction is defined as follows:

\begin{equation}
	V^{1}(r) = - \frac{C_{F}C_{A}}{4} \frac{\alpha_{s}^{2}}{r^{2}},
\end{equation}

where the Casimir charges $C_{F}$ and $C_{A}$ correspond to the fundamental and adjoint representations, with values of $\frac{4}{3}$ and $3$, respectively \cite{Koma:2006si}. Furthermore, the incorporation of spin-dependent terms into the model has been approached perturbatively, as discussed in \cite{Lucha:1991vn}. This method enhances our comprehension of the energy splittings between orbital and radial excitations across various states, shedding light on the dynamics and structure of hadronic systems. By including spin-dependent effects within the theoretical framework, a more detailed and accurate description of the mass spectra and state properties is achieved, facilitating deeper insights into the underlying physics of quark interactions and confinement phenomena.

The three spin-dependent interactions utilized in this study draw inspiration from the Breit-Fermi Hamiltonian for one-gluon exchange, as discussed in references \cite{Lucha:1991vn,Voloshin:2007dx}. By employing first-order perturbation theory to incorporate these interactions as energy corrections, the spin-dependent terms are calculated. Specifically, the tensor interaction potential $V_{T}$ and spin-orbit interaction potential $V_{LS}$ contribute to describing the fine structure of the states, elucidating the intricate details of their orbital and spin characteristics. Additionally, the spin-spin interaction potential $V_{SS}$ is responsible for delineating the hyperfine splitting observed within these states, highlighting the influence of quark spin dynamics on the overall energy spectra. This approach enhances our understanding of the internal structure and properties of hadronic systems, bridging theoretical predictions with experimental observations through the comprehensive analysis of spin-dependent interactions.

\begin{equation}
	\begin{split}
		V_{SD}(r)  &= V_{T}(r) + V_{LS}(r) + V_{SS}(r). \\
		&= \biggl( - \frac{k_{s}\alpha_{s}}{4} \frac{12\pi}{M_{\mathcal{D}}M_{\bar{\mathcal{D}}}}\frac{1}{r^{3}}\biggl) \; \biggl(-\frac{1}{3}(S_{1}\cdotp S_{2}) + \frac{(S_{1} \cdotp r) {(S_{2}\cdotp r)}}{r^{2}}\biggl) \\
		& + \biggl(-\frac{3\pi k_{s}\alpha_{s}}{2M_{\mathcal{D}}M_{\bar{\mathcal{D}}}}\frac{1}{r^{3}}  -  \frac{b}{2M_{\mathcal{D}}M_{\bar{\mathcal{D}}}}\frac{1}{r}   \biggl)(L\cdotp S) \\
		&+ \biggl(- \frac{k_{s}\alpha_{s}}{3} \frac{8\pi}{M_{\mathcal{D}}M_{\bar{\mathcal{D}}}} \frac{\sigma}{\sqrt{\pi}}^{3} exp^{-\sigma^{2}r^{2}}\biggl) (S_{1}\cdotp S_{2})
	\end{split}
\end{equation} 

where the masses of quark and antiquark are represented by $M_{\mathcal{D}}$ and $M_{\bar{\mathcal{D}}}$ in the case of meson. Similarly, they represent the masses of diquark and antidiquark in the case of tetraquark. The solution of diagonal matrix elements of spin-$\frac{1}{2}$ and spin-1 particles determines the value of $(S_{1}\cdotp S_{2})$ \cite{Debastiani:2017msn}. $\sigma$ is a parameter that has been introduced as a replacement for the Dirac Delta function and $k_{s}$ is the color factor associated with the state.

The term $( - \frac{k_{s}\alpha_{s}}{4} \frac{12\pi}{M_{\mathcal{D}}M_{\bar{\mathcal{D}}}}\frac{1}{r^{3}}) \; (-\frac{1}{3}(S_{1}\cdotp S_{2}) + \frac{(S_{1} \cdotp r) {(S_{2}\cdotp r)}}{r^{2}})$ in $V_{SD}(r)$ is known as the tensor interaction potential. Similarly, the term $(-\frac{3\pi k_{s}\alpha_{s}}{2M_{\mathcal{D}}M_{\bar{\mathcal{D}}}}\frac{1}{r^{3}}  -  \frac{b}{2M_{\mathcal{D}}M_{\bar{\mathcal{D}}}}\frac{1}{r})(L\cdotp S)$ is known as the spin-orbit interaction potential. The term, $(- \frac{k_{s}\alpha_{s}}{3} \frac{8\pi}{M_{\mathcal{D}}M_{\bar{\mathcal{D}}}} \frac{\sigma}{\sqrt{\pi}}^{3} exp^{-\sigma^{2}r^{2}}) (S_{1}\cdotp S_{2})$ is known as the spin-spin interaction potential. These spin-dependent interaction potentials have been discussed in greater detail in our previous work \cite{Tiwari:2022azj,Lodha:2024poy}. Ref. \cite{Tiwari:2021iqu,Debastiani:2017msn,Lundhammar:2020xvw} illustrates this term and its functioning in further details. 

$J^{PC}$ values of different tetraquark states have been predicted by using the relations $P_{T} = (-1)^{L_{T}}$ and $C_{T} = (-1)^{L_{T}+S_{T}}$, where $S_{T}$ and $L_{T}$ are total spin and total angular momentum, respectively \cite{Lucha:1995zv}.     

\section{Spectroscopy}
\subsection{Meson Spectra}

Inspired by our previous studies, we begin by calculating the mass spectra of pions and obtain the fitting parameters for diquarks and tetraquarks from these spectra. The fitting parameters $(M_{q},\alpha_{s},b,\sigma)$ are calculated for the present work with values $M_{q}=0.37$ $GeV$, $\alpha_{s}=0.95$, $b=0.1325$ $GeV^{2}$, $\sigma=0.83$ $GeV$ and $V_{0}=-0.47$ $GeV$ for pions ($q\bar{q}$). 

SU(3) color symmetry only permits colorless quark combinations $\ket{q\bar{q}}$ to form a color singlet state. With a color factor, $k_{s} = -\frac{4}{3} $, pions have representation exhibits $\bar{\textbf{3}}\otimes{\textbf{3}} = \textbf{1} \oplus \textbf{8}$ \cite{Debastiani:2017msn}. The masses of each pion $[q\bar{q}]$ state are given by,
\begin{equation}
	M_{(q\bar{q})} =  M_{q} + M_{\bar{q}} + E_{(q\bar{q})} + \braket{V^{1}(r)}.
\end{equation} 
where $M_{q}$ and $M_{\bar{q}}$ are constituent masses of up/down qurak and up/down anti-quark, respectively. The final mass derived from this calculation includes contributions from spin-dependent terms and relativistic mass corrections. The masses generated in the present work are compared with the experimental data from the most recent updated Particle Data Group \cite{ParticleDataGroup:2022pth} and show good agreement with it. Table \ref{SWaveMesonmass} includes the computed mass for pions as well as comparisons between several theoretical models and PDG for the S, P and D waves. Similarly, G and F waves are tabulated in table \ref{GWaveMesonmass}.

\begin{center}
	\begin{table*}
		\caption{Pion mass spectra and comparison with PDG \cite{ParticleDataGroup:2022pth} and various theoretical models for S, P and D wave in MeV}
		\label{SWaveMesonmass}
		\begin{tabular*}{\textwidth}{@{\extracolsep{\fill}}lrrrrrrrrrrl@{}}
			\hline

			State & \multicolumn{1}{c}{$J^{PC}$} & \multicolumn{1}{c}{Meson} & \multicolumn{1}{c}{Mass$_{NR}$} & \multicolumn{1}{c}{Mass$_{SR}$} & \multicolumn{1}{c}{\cite{ParticleDataGroup:2022pth}}    &\cite{Godfrey:1985xj}   &\cite{Ishida:1986vn} &\cite{Vijande:2004he} &\cite{Ebert:2009ub} &\cite{Kojo:2022psi} &\cite{Fischer:2014xha}\\
			\hline
			&  &  &  &  &&&&&&&\\
			$1 ^{1}S_{0}$ &{$0^{-+}$} &$\pi^{\pm}$ &139.60 &139.49 &139.57&150&140&139&154&160&138\\
			$2 ^{1}S_{0}$ &{$0^{-+}$} & &1149.95 &1143.00 & &1300&1270&1288&1292&1280&1103\\
			$3 ^{1}S_{0}$ &{$0^{-+}$} &$\pi(1800)$ &1829.05 &1780.38  &$1810^{+9}_{-11}$  &1880&1850&1720&1788&1820&1770\\
			$4 ^{1}S_{0}$ &{$0^{-+}$} &$\pi(2360)$ &2386.67 &2331.73  &$2360\pm25$  &&2150&&2073&2220&\\
			$5 ^{1}S_{0}$ &{$0^{-+}$} & &2879.44 &2902.35  &  &&&&2385&&\\
			&&&&&&&&&&&\\
			$1 ^{3}S_{1}$ &$1^{--}$ &$\rho(770)$ &772.81 &774.76   &$775.11\pm0.34$&770&763&772&776&760&757\\
			$2 ^{3}S_{1}$ &$1^{--}$ &$\rho(1570)$ &1521.00 &1514.19  &$1570\pm70$&1450&1600&1478&1486&1440&1023\\
			$3 ^{3}S_{1}$ &$1^{--}$ & &2118.96 &2064.00  &&2000&2140&1802&1921&1870&1332\\
			$4 ^{3}S_{1}$ &$1^{--}$ & &2632.29 &2520.48  &&&2440&1927&2195&2220&\\
			$5 ^{3}S_{1}$ &$1^{--}$ & &3095.06 &2951.95  &&&&&2491&&\\
			
			&  &  &  & &&& &&&&\\
			
			\hline
			
			&  &  &  &  &&&&&&&\\
			
			$1 ^{1}P_{1}$ &$1^{+-}$ &$b_{1}(1235)$ &1261.67  &1278.54  &$1229.5\pm3.2$&1220&1280&1234&1258&1270&852\\
			$2 ^{1}P_{1}$ &$1^{+-}$ & &1881.24  &1844.86  &&1780&1910&&1721&1700&1017\\
			$3 ^{1}P_{1}$ &$1^{+-}$ & &2416.74  &2323.40  &&&2390&&2007&2060&1352\\
			$4 ^{1}P_{1}$ &$1^{+-}$ & &2896.31  &2773.45  &&&&&2264&&\\
			$5 ^{1}P_{1}$ &$1^{+-}$ & &3337.14  &3265.39  &&&&&&&\\
			&&&&&&&&&&&\\
			$1 ^{3}P_{0}$ &$0^{++}$ & &860.81  &903.32  &&1090&940&&1176&&643\\
			$2 ^{3}P_{0}$ &$0^{++}$ &$a_{0}(1450)$ &1497.74  &1503.06  &$1439\pm34$&1780&1660&&1679&&1267\\
			$3 ^{3}P_{0}$ &$0^{++}$ &$a_{0}(2020)$ &2042.27  &1993.10  &$2025\pm30$&&2160&&1993&&1769\\
			$4 ^{3}P_{0}$ &$0^{++}$ & &2526.22  &2434.91  &&&&&2250&&\\
			$5 ^{3}P_{0}$ &$0^{++}$ & &2969.96  &2891.77  &&&&&&&\\
			&&&&&&&&&&&\\
			$1 ^{3}P_{1}$ &$1^{++}$ &$a_{1}(1260)$ &1298.69  &1316.00  &$1230\pm40$&1240&1240&1205&1254&1190&969\\
			$2 ^{3}P_{1}$ &$1^{++}$ &$a_{1}(1930)$ &1926.84  &1896.52  &$1930^{+30}_{-70}$&1820&1900&1677&1742&1680&1188\\
			$3 ^{3}P_{1}$ &$1^{++}$ & &2469.25  &2379.65  &&&2380&&2039&2030&\\
			$4 ^{3}P_{1}$ &$1^{++}$ & &2951.46  &2818.06  &&&&&2286&&\\
			$5 ^{3}P_{1}$ &$1^{++}$ & &3394.24  &3272.04  &&&&&&&\\
			
			&&&&&&&&&&&\\
			$1 ^{3}P_{2}$ &$2^{++}$ &$a_{2}(1320)$ &1333.58 &1369.96  &$1318.2\pm0.6$&1310&1300&1327&1317&1330&1154\\
			$2 ^{3}P_{2}$ &$2^{++}$ &$a_{2}(1990)$ &1981.11 &1962.23  &$2003\pm10\pm19$&1820&1950&1732&1779&1740&\\
			$3 ^{3}P_{2}$ &$2^{++}$ & &2534.24 &2450.39  &&&2440&&2048&2070&\\
			$4 ^{3}P_{2}$ &$2^{++}$ & &3023.49 &2891.66  &&&&&2297&&\\
			$5 ^{3}P_{2}$ &$2^{++}$ & &3471.33 &3348.41  &&&&&&&\\
			
			&  &  &  &  &&&&&&&\\
			\hline
			&  &  &  &  &&&&&&&\\
			
			$1 ^{1}D_{2}$&$2^{-+}$  &$\pi_{2}(1670)$  &1624.53  &1625.93 &$1670.6^{+2.9}_{-1.2}$&1680&1690&1600&1643&1620&1227\\
			$2 ^{1}D_{2}$ &$2^{-+}$ &$\pi_{2}(2100)$  &2178.51  &2123.29 &$2090\pm29$&2130&2230&&1960&1980&\\
			$3 ^{1}D_{2}$ &$2^{-+}$ &  &2676.14  &2567.08  &&&&&2216&2280&\\
			$4 ^{1}D_{2}$ &$2^{-+}$ &  &3130.30  &3005.47  &&&&&&&\\
			$5 ^{1}D_{2}$ &$2^{-+}$ &  &3551.86  &3506.44  &&&&&&&\\
			&&&&&&&&&&&\\
			$1 ^{3}D_{1}$ &$1^{--}$ &$\rho(1570)$  &1571.39  &1581.48  &$1570\pm70$&1660&1600&&1557&1580&\\
			$2 ^{3}D_{1}$ &$1^{--}$ &$\rho(2150)$  &2128.68  &2083.06  &&2150&2150&1826&1895&1950&\\
			$3 ^{3}D_{1}$ &$1^{--}$ &  &2629.68  &2528.99  &&&&&2168&2210&\\
			$4 ^{3}D_{1}$ &$1^{--}$ &  &3085.83  &2966.02  &&&&&&&\\
			$5 ^{3}D_{1}$ &$1^{--}$ &  &3509.73  &3460.00  &&&&&&&\\
			&&&&&&&&&&&\\
			$1 ^{3}D_{2}$ &$2^{--}$  &&1606.43  &1626.54  &&1700&1670&&1661&1610&1202\\
			$2 ^{3}D_{2}$ &$2^{--}$ &&2175.93  &2135.54 &&2150&2220&&1983&1970&\\
			$3 ^{3}D_{2}$ &$2^{--}$ &&2685.45  &2586.48  &&&&&2241&2270&\\
			$4 ^{3}D_{2}$ &$2^{--}$ &&3145.95  &3027.84  &&&&&&&\\
			$5 ^{3}D_{2}$ &$2^{--}$ &&3575.01  &3524.48  &&&&&&&\\
			&&&&&&&&&&&\\
			$1 ^{3}D_{3}$ &$3^{--}$  &$\rho_{3}(1690)$&1552.85  &1644.00  &$1688.8\pm2.1$&1680&1690&1636&1714&1630&1528\\
			$2 ^{3}D_{3}$ &$3^{--}$  &$X(2110)$&2137.05  &2189.28  &$2110\pm10$&2130&2250&1878&2066&1990&\\
			$3 ^{3}D_{3}$ &$3^{--}$  &&2656.00  &2673.88  &&&&&2309&2290&\\
			$4 ^{3}D_{3}$ &$3^{--}$  &&3124.85  &3112.25  &&&&&&&\\
			$5 ^{3}D_{3}$ &$3^{--}$  &&3557.75  &3517.33  &&&&&&&\\
			\hline
			
		\end{tabular*}
	\end{table*}
\end{center}

\begin{center}
	\begin{table*}
		\caption{Pion mass spectra and comparison with PDG \cite{ParticleDataGroup:2022pth} and various theoretical models for F and G wave  in MeV}
		\label{GWaveMesonmass}
		\begin{tabular*}{\textwidth}{@{\extracolsep{\fill}}lrrrrrrrl@{}}
			\hline

			State & \multicolumn{1}{c}{$J^{P}$} & \multicolumn{1}{c}{Meson} & \multicolumn{1}{c}{Mass$_{NR}$} & \multicolumn{1}{c}{Mass$_{SR}$} & \multicolumn{1}{c}{\cite{ParticleDataGroup:2022pth}}    &\cite{Godfrey:1985xj}   &\cite{Ishida:1986vn}  &\cite{Ebert:2009ub} \\
			\hline
			&  &  &   &&&&&\\
			
			$1 ^{1}F_{3}$&$3^{+-}$&&1953.28  &1929.47  &&2030&2000&1884\\
			$2 ^{1}F_{3}$&$3^{+-}$&&2461.46  &2380.39  &&&&2164\\
			$3 ^{1}F_{3}$&$3^{+-}$&&2930.52  &2807.75  &&&&\\
			$4 ^{1}F_{3}$&$3^{+-}$&&3363.46  &3261.79  &&&&\\
			
			&&&&&&&&\\
			$1 ^{3}F_{2}$&$2^{++}$&$X(1870)$&1888.17  &1884.54  &$1870\pm40$&2050&1970&1797\\
			$2 ^{3}F_{2}$&$2^{++}$&&2406.13  &2341.80  &&&&2091\\
			$3 ^{3}F_{2}$&$2^{++}$&&2881.91  &2773.58  &&&&\\
			$4 ^{3}F_{2}$&$2^{++}$&&3320.16  &3231.33  &&&&\\
			
			&&&&&&&&\\
			$1 ^{3}F_{3}$&$3^{++}$&$a_{3}(1875)$&1868.47  &1877.89  &$1874\pm43\pm96$&2050&1990&1910\\
			$2 ^{3}F_{3}$&$3^{++}$&$a_{3}(2275)$&2395.22  &2342.22  &$2275\pm35$&&&2191\\
			$3 ^{3}F_{3}$&$3^{++}$&&2877.90  &2778.81  &&&&\\
			$4 ^{3}F_{3}$&$3^{++}$&&3321.16  &3239.41  &&&&\\
			
			&&&&&&&&\\
			$1 ^{3}F_{4}$&$4^{++}$&$a_{4}(1970)$&1808.49  &2010.40  &$1967\pm16$&2010&2000&2018\\
			$2 ^{3}F_{4}$&$4^{++}$&$X(2360)$&2345.44  &2510.09  &$2360\pm10$&&&2284\\
			$3 ^{3}F_{4}$&$4^{++}$&&2834.55  &2965.42  &&&&\\
			$4 ^{3}F_{4}$&$4^{++}$&&3282.82  &3384.18  &&&&\\
			\hline
			&  &  &  &  &&&&\\
			
			$1 ^{1}G_{4}$&$4^{-+}$&$\pi_{4}(2250)$&2257.23  &2201.72  &&2330&&2092\\
			$2 ^{1}G_{4}$&$4^{-+}$&&2731.47  &2623.97  &&&&2344\\
			$3 ^{1}G_{4}$&$4^{-+}$&&3175.08  &3052.80  &&&&\\
			
			&&&&&&&&\\
			$1 ^{3}G_{3}$&$3^{--}$&$X(2110)$&2151.53  &2128.44  &$2110 \pm10$&2370&&2002\\
			$2 ^{3}G_{3}$&$3^{--}$&&2639.36  &2560.74  &&&&2667\\
			$3 ^{3}G_{3}$&$3^{--}$&&3093.04  &2997.05  &&&&\\
			
			&&&&&&&&\\
			$1 ^{3}G_{4}$&$4^{--}$&&2117.39  &2107.97  &&2340&&2122\\
			$2 ^{3}G_{4}$&$4^{--}$&&2611.95  &2545.13  &&&&2375\\
			$3 ^{3}G_{4}$&$4^{--}$&&3070.65  &2985.06  &&&&\\
			
			&&&&&&&&\\
			$1 ^{3}G_{5}$&$5^{--}$&$\rho_{5}(2350)$&2070.62 &2425.17 &$2330\pm35$&2300&&2264\\
			$2 ^{3}G_{5}$&$5^{--}$&&2572.02 &2891.69 &&&&2472\\
			$3 ^{3}G_{5}$&$5^{--}$&&3035.82 &3323.92 &&&&\\
			\hline
		\end{tabular*}
	\end{table*}
\end{center}		

\subsection{Diquarks}
Diquarks $(\mathcal{D})$ are hypothesized bound states of two quarks interacting with each other through gluonic exchange and are considered to be the essential constituents shaping baryons. Unlike their constituent quarks, diquarks are not observed as independent entities and the quark-diquark model suggests their existence within baryons. The implications of diquarks extend beyond the realm of baryon structure. They could serve as the foundational building blocks of exotic hadrons like tetraquarks and pentaquarks. Furthermore, unraveling the intricacies of diquark interactions could provide us with a deeper understanding of the strong force, the fundamental interaction that governs the behavior of quarks. The antiparticles of diquarks are known as anti-diquarks ($\mathcal{\bar{D}}$). 

When diquarks and anti-diquarks interact, they form composite systems rather than behaving as point-like entities. Due to Pauli's exclusion principle, the ground-state wavefunction of these systems exhibits an antisymmetric nature. Diquarks can exhibit different natures, such as purely scalar, axial-vector, or vector. Scalar diquarks are often referred to as "good diquarks," whereas axial-vector diquarks are termed "bad diquarks," as discussed in detail in \cite{Esposito:2016noz}.

In QCD color symmetry, the fundamental representation for a diquark is expressed as $\textbf{3}\otimes\textbf{3}=\bar{\textbf{3}}\oplus \textbf{6}$, as outlined in \cite{Debastiani:2017msn}. Similarly, an anti-diquark in the $\bar{\textbf{3}}$ representation is defined by $\bar{\textbf{3}}\otimes\bar{\textbf{3}}=\textbf{3}\oplus\bar{\textbf{6}}$. Since a tetraquark is generally considered a four-body problem, the treatment of tetraquark as a bound of the state of diquark and antidiquark reduces it to a two-body problem \cite{Fredriksson:1981mh}. The $\bar{\textbf{3}}\otimes{\textbf{3}}$  color coupling leads to the formation of the $\textbf{1}\otimes{\textbf{1}}$ state and the $\textbf{8}\otimes\bar{\textbf{8}}$ state. As the color factor $k_{s}$ in QCD color symmetry for triplet-antitriplet state is $-\frac{2}{3}$, the short distance part, $\frac{1}{r}$ of the interaction has an attractive nature \cite{Deba}. 
The methodology employed to calculate the masses of diquarks and anti-diquarks mirrors that used for pions. The masses of diquarks and anti-diquarks studied in the present work are given by,    

\begin{equation}
	M_{(qq)} = 2M_{q} + E_{(qq)} + \braket{V^{1}(r)}\\
\end{equation} 
where $M_{q}$ is the mass of constituent quarks in the diquark. 

\begin{table*}	
	\centering
	\caption{Mass spectra and comparison  for various qq diquarks/ anti-diquarks. A comparison with various theoretical models is made. All units are in MeV.}
	\label{diquark}
	\begin{tabular}{ccccccccc}
		
		\hline
		\multicolumn{2}{c}{ Mass$_{SR}$} &\multicolumn{2}{c}{Mass$_{NR}$} & \multirow{2}{*}{\cite{Maris:2002yu}} &\multirow{2}{*}{\cite{Faustov:2021hjs}}  &\multirow{2}{*}{\cite{Chen:2023ngj}} &\multirow{2}{*}{\cite{Ferretti:2019zyh}}
		&\multirow{2}{*}{\cite{Yin:2021uom}} \\ 
		Triplet &Sextet &Triplet &Sextet&&&&&\\
		\hline
		942 &1126 &989&1130  &1020  &909     &970 &840&1060
		\\
		\hline
		
	\end{tabular}	
\end{table*}

\subsection{Tetraquark Spectra}
A tetraquark in a color singlet state can arise from two distinct diquark-antidiquark pairings: (i) a color anti-triplet diquark paired with a color triplet anti-diquark, or (ii) a color sextet diquark paired with a color anti-sextet anti-diquark. For the color singlet state combinations $\bar{\textbf{3}}-\textbf{3}$ and $\textbf{6}-\bar{\textbf{6}}$, the corresponding color factors are denoted as $k_{s}$, with values of $-\frac{4}{3}$ and $\frac{-10}{3}$, respectively, as discussed in \cite{Deba,Debastiani:2017msn,Lodha:2024erj}. Through the combination of spin-1 diquarks and anti-diquarks, a color singlet tetraquark can be formed, represented as $\ket{QQ|^{3}\otimes|\bar{Q}\bar{Q}|^{\bar{3}}} = \textbf{1}\oplus\textbf{8}$.

A tetraquark with four light quarks can have more than one internal structure based on the quark flavor of the diquarks or anti-diquarks involved. However, since the mass difference between up quark and down quark constituent mass is merely a few MeVs, this distinction is very difficult to make. If the constituent quark content of the given tetraquark consists of either only up, only down quark or a combination of both, the tetraquark state behaves as quarkonia. Hence, the mass spectra for $T_{4q}$ have the formulation,

\begin{equation}
	M_{qqqq} =  M_{qq} + M_{\bar{q}\bar{q}} + E_{(qq\bar{q}\bar{q})} + \braket{V^{1}(r)}\\		
\end{equation}

The mass spectra of calculated tetraquark states are influenced by various components, including the Cornell-like potential $V^{0}_{C+L}$, the relativistic term represented by $\braket{V^{1}(r)}$, and spin-dependent contributions. These spin-dependent contributions are computed separately to ascertain their individual impacts. By combining the total spin $S_{T}$ with the orbital angular momentum $L_{T}$, denoted as $S_{T} \otimes L_{T}$, the color singlet state of the tetraquark is derived. This process involves coupling the total spin and orbital angular momentum to determine the appropriate quantum states corresponding to the tetraquark's color singlet configuration. 
The interaction of these elements, including the interplay between spin-dependent effects and orbital angular momentum, is critical to understanding the overall structure and characteristics of tetraquark states within the framework of quantum chromodynamics (QCD). This approach enables a detailed analysis of how different physical parameters contribute to the observed mass spectra and properties of tetraquark configurations, shedding light on the underlying dynamics governing these exotic hadrons.

\begin{equation}
	\ket{T} = \ket{S_{d},S_{\bar{d}},S_{T},L_{T}}_{J_{T}},
\end{equation}

where the spins of diquark and anti-diquark are denoted by  $S_{d}$ and $S_{\bar{d}}$, respectively. For mesons, diquarks, and anti-diquarks, only two spin combinations were viable. However, in the case of tetraquarks formed from spin-1 diquarks and anti-diquarks, three distinct spin combinations become feasible. This expanded set of spin possibilities reflects the increased complexity and diverse configurations achievable within tetraquark structures, highlighting the intricacy of quark interactions and multi-quark states.

\begin{equation}
	\begin{split}
		\ket{0^{++}}_{T}=\ket{S_{d}=1,S_{\bar{d}}=1,S_{T}=0,L_{T}=0}_{J_{T=0}};\\
		\ket{1^{+-}}_{T}=\ket{S_{d}=1,S_{\bar{d}}=1,S_{T}=1,L_{T}=0}_{J_{T=1}};\\
		\ket{2^{++}}_{T}=\ket{S_{d}=1,S_{\bar{d}}=1,S_{T}=2,L_{T}=0}_{J_{T=2}};
	\end{split}
\end{equation}

 
By utilizing one-gluon exchange interactions between a quark from a diquark and an anti-quark from an anti-diquark, it becomes possible to mix states involving ${\textbf{6}}\otimes\bar{\textbf{6}}$ and $\bar{\textbf{3}}\otimes{\textbf{3}}$. However, achieving this mixing requires a four-body approach to adequately address the tetraquark configuration. The current methodology adopts a diquark-antidiquark approximation, treating the tetraquark as a simplified two-body problem, which precludes the formation of mixed-state tetraquarks. A more comprehensive explanation of this limitation is detailed in the referenced study, \cite{Debastiani:2017msn}. The mass spectra of $\bar{\textbf{3}}\otimes{\textbf{3}}$ and ${\textbf{6}}\otimes\bar{\textbf{6}}$ states for S wave and P wave for all light tetraquark are done in table \ref{Swavetriplet}, along with comparison with other theoretical works and the two meson threshold. The parameter sensitivity of the current theoretical model for mass spectra has been discussed in great detail in our previous work. 

\begin{table}
	\centering
	\caption{Mass spectra of Tetraquarks $T_{4q}$ with $\bar{3}\otimes{3}$ and ${6}\otimes\bar{6}$ diquark-antidiquark configurations. All units are in MeV.}
	\label{Swavetriplet}
	\begin{tabular}{cccccc}
		
		\hline
		\multirow{2}{*}{State} &\multirow{2}{*}{$J^{PC}$} & \multicolumn{2}{c}{$\bar{\textbf{3}}-{\textbf{3}}$} & \multicolumn{2}{c}{${\textbf{6}}-\bar{\textbf{6}}$}\\
		&&Mass$_{NR}$ &Mass$_{SR}$& Mass$_{NR}$ & Mass$_{SR}$ \\

		\hline
		&&&&&\\
		$1 ^{1}S_{0}$ &	$0^{++}$ &1206.33	&1044.40 &-&-\\
		$2 ^{1}S_{0}$&$0^{++}$ 	&2317.20 &2269.12 &2621.95	&2532.22\\
		$3 ^{1}S_{0}$&$0^{++}$ 	&2862.51 &2811.55 &3678.35	&3584.39\\
		&&&&&\\
		$1 ^{3}S_{1}$ &$1^{+-}$	&1474.34	&1360.58&&\\
		$2 ^{3}S_{1}$&$1^{+-}$	&2403.44	&2377.99&&\\
		$3 ^{3}S_{1}$&$1^{+-}$	&2919.07	&2882.44&&\\
		&&&&&\\
		$1 ^{5}S_{2}$ &	{$2^{++}$}	&2013.02	&2040.25&&\\
		$2 ^{5}S_{2}$&{$2^{++}$} 	&2578.17	&2590.14&&\\
		$3 ^{5}S_{2}$&{$2^{++}$} 	&3033.53	&3023.61&&\\
		\hline
		
		&&&&&\\
		$1 ^{1}P_{1}$&	$1^{--}$&2297.82	&2296.75 &2436.58	&2347.23\\
		$2 ^{1}P_{1}$&$1^{--}$ 	&2787.46	&2771.03&3466.74	&3388.26  \\
		$3 ^{1}P_{1}$&$1^{--}$	&3195.36	&3156.18	&4258.62	&4160.75 \\
		&&&&&\\
		$1 ^{3}P_{0}$ &	$0^{-+}$	&1898.38	&1890.94 &-	&-\\
		$2 ^{3}P_{0}$&$0^{-+}$ 	&2447.26	&2429.52 &1813.00	&1750.10\\
		$3 ^{3}P_{0}$&$0^{-+}$	&2877.23	&2840.24 &2766.63	&2691.91\\
		&&&&&\\
		$1 ^{3}P_{1}$ &	$1^{-+}$	&2305.39	&2307.74 &2430.86	&2355.23 \\
		$2 ^{3}P_{1}$&$1^{-+}$	&2793.41	&2781.35 	&3437.23	&3368.24 \\
		$3 ^{3}P_{1}$&$1^{-+}$	&3198.31	&3164.37 	&4218.51	&4119.70  \\
		&&&&&\\
		$1 ^{3}P_{2}$&$2^{-+}$	&2429.75	&2437.83 &3129.93	&3087.11 \\
		$2 ^{3}P_{2}$&$2^{-+}$	&2904.66	&2895.98 	&3999.76	&3936.83  \\
		$3 ^{3}P_{2}$&$2^{-+}$	&3303.37	&3271.75 	&4728.20	&4628.76 \\
		&&&&&\\
		$1 ^{5}P_{1}$ &	$1^{--}$ &1895.31	&1896.38&&\\
		$2 ^{5}P_{1}$&$1^{--}$	&2446.61	&2439.68&&\\
		$3 ^{5}P_{1}$&$1^{--}$	&2875.61	&2849.00&&\\
		&&&&&\\
		$1 ^{5}P_{2}$ &$2^{--}$	&2404.72	&2417.19&&\\
		$2 ^{5}P_{2}$&$2^{--}$	&2883.65	&2881.05&&\\
		$3 ^{5}P_{2}$&$2^{--}$	&3280.50	&3256.73&&\\
		&&&&&\\
		$1 ^{5}P_{3}$&	$3^{--}$&2587.96	&2611.87&&\\
		$2 ^{5}P_{3}$&$3^{--}$	&3046.36	&3052.23&&\\
		$3 ^{5}P_{3}$&$3^{--}$	&3436.22	&3417.47&&\\
		\hline
	\end{tabular}
\end{table}

\begin{table}
	\centering
	\caption{Two meson threshold for different tetraquark states}
	\label{twomesonthreshold}
	\begin{tabular}{cccc}
		\hline	
		\multirow{2}{*}{State} & \multirow{2}{15mm}{Two-meson Threshold} 
		&\multicolumn{2}{c}{Threshold Mass}\\
		&&Semi-relativistic&Non-relativistic\\   
		\hline
		$^{1}S_{0}$ & $\pi(1S)\pi(1S)$ &273.20 &274.98 \\
		$^{3}S_{1}$ & $\pi(1S)\rho(1S)$ &909.41 &912.25 \\
		$^{5}S_{2}$ & $\rho(1S)\rho(1S)$ &1545.62 &1549.52 \\
		$^{3}P_{0}$ & $\pi(1S)$  $a_{0}(1P)$ &997.41 &1040.81 \\
		$^{3}P_{1}$ & $\pi(1S)$  $a_{1}(1P)$ &1435.29 &1453.49 \\
		$^{3}P_{2}$ & $\pi(1S)$  $a_{2}(1P)$ &1470.18 &1507.45 \\
		$^{5}P_{1}$ & $\pi(1S)$  $b_{1}(1P)$ &1398.27 &1416.03 \\
		$^{5}P_{2}$ & $\rho(1S)$  $a_{1}(1P)$ &2071.50 &2090.76 \\
		$^{5}P_{3}$ & $\rho(1S)$  $a_{2}(1P)$ &2106.39 &2144.72 \\
		\hline
	\end{tabular}
\end{table}

\begin{table*}
	\centering
	\caption{Tetraquark assignment for various resonances }
	\label{Assignment}
	\begin{tabular}{ccccc}
		\hline	
		State &Resonance &Mass$_{\text{Cal.}}$  &Mass$_{\text{Exp.}}$ &$\Gamma$ \\
		\hline
		$1 ^{1}S_{0_{\bar{\textbf{3}}-{\textbf{3}}}}$&	$a_{0}(980)$&1044.40&$980\pm20$&50-100\\
		$1 ^{1}S_{0_{\bar{\textbf{3}}-{\textbf{3}}}}$&	$f_{0}(980)$&1044.40&$990\pm20$&10-100\\
		$1 ^{1}S_{0_{\bar{\textbf{3}}-{\textbf{3}}}}$&	$X(1070)$&1044.40&$1072\pm1$&$3.5\pm0.5$\\
		$2 ^{1}S_{0_{{\textbf{6}}-\bar{\textbf{6}}}}$&	$X(2540)$&2532.22&$2539\pm14^{+38}_{-14}$&$274^{+77+126}_{-61-123}$\\
		$2 ^{1}S_{0_{\bar{\textbf{3}}-{\textbf{3}}}}$&	$f_{0}(2330)$&2317.20&2312-2419&65-274\\
		$1 ^{5}S_{2_{\bar{\textbf{3}}-{\textbf{3}}}}$&	$a_{2}(1990)$&2013.02&$2003\pm10\pm19$&$249\pm23\pm32$\\
		$1 ^{5}S_{2_{\bar{\textbf{3}}-{\textbf{3}}}}$&	$f_{2}(2000)$&2013.02&$2001\pm10$&$312\pm34$\\
		$1 ^{5}S_{2_{\bar{\textbf{3}}-{\textbf{3}}}}$&	$f_{2}(2010)$&2013.02&$2011^{+60}_{-80}$&$202\pm60$\\
		$1 ^{5}S_{2_{\bar{\textbf{3}}-{\textbf{3}}}}$&	$a_{2}(2030)$&2040.25&$2030\pm20$&$205\pm30$\\
		$1 ^{1}P_{1_{\bar{\textbf{3}}-{\textbf{3}}}}$&	$\omega(2290)$&2297.82&$2290\pm20$&$275\pm35$\\
		$1 ^{1}P_{1_{{\textbf{6}}-\bar{\textbf{6}}}}$&	$\omega(2330)$&2347.23&$2330\pm30$&$435\pm75$\\
		\hline
	\end{tabular}
\end{table*}

\section{Decay}

In the realm of hadrons, decay widths represent essential parameters, elucidating the lifetimes and decay mechanisms across various channels. These widths precisely gauge the likelihood of a particle transitioning into specific final states, offering insights into the intricate interplay of strong and weak interactions governing hadron decay. Expressed as $\Gamma$, decay width quantifies the probability of decay within a given timeframe. A larger $\Gamma$ indicates faster decay, while a smaller one suggests a longer-lived particle. Through meticulous analysis of decay widths, physicists delve into the fundamental forces and dynamics within the quark-gluon framework, unraveling the complex nature of hadronic matter. This introduction lays the foundation for exploring the pivotal role of decay widths in understanding the properties and behaviors of hadrons. 

Studying the decay widths of bound states provides critical insights into their internal structures. Investigating tetraquark decay dynamics, given its complex four-body nature, is particularly challenging. This work explores the decay properties of all light tetraquark states. Inspired by references \cite{Becchi:2020mjz,Becchi:2020uvq}, the study of tetraquarks involves re-arrangement phenomena. The diquark-antidiquark annihilation model, an extension of the quark-antiquark annihilation for mesons \cite{Devlani:2011zz}, is applied to analyze gluonic, leptonic, and photonic decay channels. Additionally, indirect tetraquark decay via leptonic, photonic, and analogous decay channels of rearranged hadrons is calculated.

\subsection{Annihilation decay}

In continuation of our previous study, the present work explores the annihilation decays of tetraquarks, treating the diquark-antidiquark pair similarly to the quark-antiquark pair in heavy quarkonia. This approach extends the formulation to include leptonic, photonic, and gluonic decay modes. We have calculated and compiled various annihilation decay rates for $T_{^{3}S_{1}}$ and $T_{^{1}S_{0}}$ tetraquark states, as presented in Table \ref{annihilationdecay}. This investigation aims to provide a comprehensive understanding of the decay properties of tetraquark states under annihilation processes.

The $^{1}S_{0}$ $qq\bar{q}\bar{q}$ tetraquark can decay through 2 channels: annihilation into 2 gluons and annihilation into 2 photons. Similarly, the $^{3}S_{1}$ $qq\bar{q}\bar{q}$ tetraquark can decay through 3 channels: annihilation into 3 gluons, 3 photons, and 2 leptons. The gluonic decay width, with first-order radiative corrections, is given by \cite{Segovia:2016xqb,Kwong:1987ak,Kwong:1988ae,Belanger:1987cg}:

\begin{equation}
	\Gamma_{ ^{1}S_{0} \; qq\bar{q}\bar{q}\rightarrow g g} = \frac{2 \alpha_{s}^{2} |R_{nl}(0)|^{2}}{3m_{qq}^{2}},
\end{equation}

\begin{equation}
	\Gamma_{^{3}S_{1} \; qq\bar{q}\bar{q}\rightarrow ggg} = \frac{10(\pi^{2}-9) \alpha_{s}^{3} |R_{nl}(0)|^{2}}{81\pi m_{qq}^{2}},
\end{equation}
where $|R_{nl}(0)|^{2}$ is the square modulus of the radial wave function at the origin. $|R_{nl}(0)|^{2}$ is given by \cite{Lucha:1991vn}, 

\begin{equation}
	|R_{nl}(0)|^{2}=4\pi\;|Y^{0}_{0}(\theta,\phi)R_{nl}(0)|^{2}=4\;\pi|\psi(0)|^{2}
\end{equation} 
where $|\psi(0)|^{2}$ is the square modulus of the wave function at the origin.

The leptonic decay width, is given by \cite{Wang:2019tqf} :
\begin{equation}
	\Gamma_{qq\bar{q}\bar{q}\rightarrow e^{+}e^{-}} = \frac{4\pi \alpha^{2} e_{Q}^{2}f_{qq\bar{q}\bar{q}}}{3M_{qq\bar{q}\bar{q}}}\times\biggr(1+2\frac{m_{e}^{2}}{M_{qq\bar{q}\bar{q}}^{2}}\biggr)\sqrt{1-4\frac{m_{e}^{2}}{M_{qq\bar{q}\bar{q}}^{2}}}
\end{equation}
\begin{equation}
	\Gamma_{qq\bar{q}\bar{q}\rightarrow \mu^{+}\mu^{-}} = \frac{4\pi \alpha^{2} e_{Q}^{2}f_{qq\bar{q}\bar{q}}}{3M_{qq\bar{q}\bar{q}}}\times\biggr(1+2\frac{m_{\mu}^{2}}{M_{qq\bar{q}\bar{q}}^{2}}\biggr)\sqrt{1-4\frac{m_{\mu}^{2}}{M_{qq\bar{q}\bar{q}}^{2}}}
\end{equation}

where $\alpha$ is the fine structure constant, $e_{Q}$ is the electric charge of diquark in units of electron charge, $m_{e}$ is the mass of electron, $m_{\mu}$ is the mass of muon, and $M_{qq\bar{q}\bar{q}}$ is the mass of $qq\bar{q}\bar{q}$ tetraquark.

The photonic decay width, including the first-order radiative correction, is given by \cite{Segovia:2016xqb,Kwong:1987ak}:

\begin{equation}
	\Gamma_{qq\bar{q}\bar{q}\rightarrow \gamma \gamma} = \frac{3 \alpha^{2} e_{Q}^{4}|R_{nl}(0)|^{2}}{m_{qq}^{2}},
\end{equation}

\begin{equation}
	\Gamma_{qq\bar{q}\bar{q}\rightarrow \gamma\gamma \gamma} = \frac{4(\pi^{2}-9) \alpha^{3} e_{Q}^{6}|R_{nl}(0)|^{2}}{3\pi m_{qq}^{2}},
\end{equation}
where $m_{qq}$ is the constituent mass of the diquark and $|R_{nl}(0)|^{2}$ is the square modulus of the radial wave function at the origin.

\begin{table*}
	\centering
	\caption{Annihilation decay rates for various tetraquark channels in MeV}
	\label{annihilationdecay}
	\begin{tabular}{cccc}
		\hline	
		\multirow{2}{15mm}{Internal color configuration}	&\multirow{2}{10mm}{Decay Channel} & \multirow{2}{15mm}{Semi-relativistic} & \multirow{2}{15mm}{Non-relativistic} \\ \\ \\ 
		\hline
		\multirow{7}{*}{${\bar{3}\otimes3}$}&$T_{^{3}S_{1}} \rightarrow e^{+}e^{-}$ &$6.903\times10^{3}$&$ 4.513\times10^{3}$ \\
		&$T_{^{3}S_{1}} \rightarrow\mu^{+}\mu^{-}$ &$6.889\times10^{3}$&$4.503\times10^{3}$ \\
		
		&$T_{^{1}S_{0}} \rightarrow2 \gamma$&$33.674\times10^{3}$&$16.497\times10^{3}$ \\
		&$T_{^{3}S_{1}} \rightarrow$ $3\gamma$ &67.031&16.497 \\
		&$T_{^{1}S_{0}} \rightarrow 2g$ &$254.541\times10^{6}$&$278.282\times10^{6}$ \\
		&$T_{^{3}S_{1}} \rightarrow 3g$ &$13.559\times10^{6}$&$12.402\times10^{6}$ \\
		\hline
	\end{tabular}
	
\end{table*}
\subsection{Re-arrangement Decay}

\begin{figure}[t]
	\centering
	\includegraphics[width=0.5\linewidth, height=0.29\textheight]{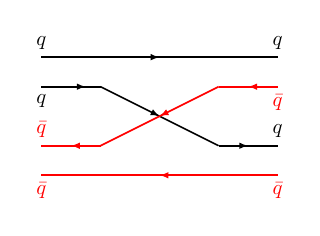}
	\caption[]{Rearrangement diagram for Tetraquark $T_{qq\bar{q}\bar{q}}$}
	
\end{figure}

The recoupling of spin wave functions is given by:

\begin{equation}		
	\biggr|{\{{(qq)}^{1}{(\bar{q}\bar{q})}^{1}\}^{0}}\biggr \rangle=-\frac{1}{2} \biggr|\bigr(q\bar{q}\bigr)^{1}\bigr(q\bar{q}\bigr)^{1}\biggr \rangle ^{0} + \frac{\sqrt{3}}{2}\biggr|\bigr(q\bar{q}\bigr)^{0}\bigr(q\bar{q}\bigr)^{0}\biggr \rangle ^{0} 	
	\label{spinwf1}		
\end{equation}

Similarly, the recoupling of color wave functions is given by:
\begin{equation}		
	|{{(qq)}_{\bar{\textbf{3}}}{(\bar{q}\bar{q})}_{\textbf{3}}}|\rangle=\sqrt{\frac{1}{3}} |\bigr(q\bar{q}\bigr)_{\textbf{1}}\bigr(q\bar{q}\bigr)_{\textbf{1}}\rangle - \sqrt{\frac{2}{3}}|\bigr(q\bar{q}\bigr)_{\textbf{8}}\bigr(q\bar{q}\bigr)_{\textbf{8}}\rangle 
	\label{colorwf33_1}		
\end{equation}

\begin{equation}		
	|{{(qq)}_{\textbf{6}}}{(\bar{q}\bar{q})}_{\bar{\textbf{6}}}|\rangle=\sqrt{\frac{2}{3}} |\bigr(q\bar{q}\bigr)_{\textbf{1}}\bigr(q\bar{q}\bigr)_{\textbf{1}}\rangle + \sqrt{\frac{1}{3}}|\bigr(q\bar{q}\bigr)_{\textbf{8}}\bigr(q\bar{q}\bigr)_{\textbf{8}}\rangle 
	\label{colorwf66_1}		
\end{equation}

By employing the Fierz transformation, different quark-antiquark pairs ($q\bar{q}$) are combined \cite{Ali:2019roi}. Through the spectator-pair method, the tetraquark decays into two mesons. The normalization of quark bilinears is set to unity. The Fierz rearrangement for various states, utilizing equations for spin wave function recoupling and color wave function recoupling, is outlined as follows:

\begin{eqnarray}
	\begin{split}
		&qq\bar{q}\bar{q}(J=0)=\biggr|\bigr(qq\bigr)^{1}_{\bar{\textbf{3}}}\bigr(\bar{q}\bar{q}\bigr)^{1}_{\textbf{3}}\biggr \rangle ^{0}_{1} \\ 
		&=-\frac{1}{2} \biggr(\sqrt{\frac{1}{3}}\biggr|\bigr(q\bar{q}\bigr)^{1}_{\textbf{1}}\bigr(q\bar{q}\bigr)^{1}_{\textbf{1}}\biggr \rangle ^{0}_{\textbf{1}}  -\sqrt{\frac{2}{3}}\biggr|\bigr(q\bar{q}\bigr)^{1}_{\textbf{8}}\bigr(q\bar{q}\bigr)^{1}_{\textbf{8}}\biggr \rangle ^{0}_{\textbf{1}}  \biggr)\\
		&+\frac{\sqrt{3}}{2} \biggr(\sqrt{\frac{1}{3}}\biggr|\bigr(q\bar{q}\bigr)^{0}_{\textbf{1}}\bigr(q\bar{q}\bigr)^{0}_{\textbf{1}}\biggr \rangle ^{0}_{\textbf{1}}  -\sqrt{\frac{2}{3}}\biggr|\bigr(q\bar{q}\bigr)^{0}_{\textbf{8}}\bigr(q\bar{q}\bigr)^{0}_{\textbf{8}}\biggr \rangle ^{0}_{1}  \biggr)
	\end{split}
	\label{eqssqq}
\end{eqnarray}

\begin{eqnarray}
	\begin{split}
		&qq\bar{q}\bar{q}(J=0)=\biggr|\bigr(qq\bigr)^{1}_{\textbf{6}}\bigr(\bar{q}\bar{q}\bigr)^{1}_{\bar{\textbf{6}}}\biggr \rangle ^{0}_{\textbf{1}} \\ 
		&=-\frac{1}{2} \biggr(\sqrt{\frac{2}{3}}\biggr|\bigr(q\bar{q}\bigr)^{1}_{\textbf{1}}\bigr(q\bar{q}\bigr)^{1}_{\textbf{1}}\biggr \rangle ^{0}_{\textbf{1}}  +\sqrt{\frac{1}{3}}\biggr|\bigr(q\bar{q}\bigr)^{1}_{\textbf{8}}\bigr(q\bar{q}\bigr)^{1}_{\textbf{8}}\biggr \rangle ^{0}_{\textbf{1}}  \biggr)\\
		&+\frac{\sqrt{3}}{2} \biggr(\sqrt{\frac{2}{3}}\biggr|\bigr(q\bar{q}\bigr)^{0}_{\textbf{1}}\bigr(q\bar{q}\bigr)^{0}_{\textbf{1}}\biggr \rangle ^{0}_{\textbf{1}}  +\sqrt{\frac{1}{3}}\biggr|\bigr(q\bar{q}\bigr)^{0}_{\textbf{8}}\bigr(q\bar{q}\bigr)^{0}_{\textbf{8}}\biggr \rangle ^{0}_{\textbf{1}}  \biggr)
	\end{split}
	\label{eq25}
\end{eqnarray}

where the color representation's dimensions are denoted using subscripts and the total spin is denoted by a superscript.

\begin{figure}[t]
	\centering
	\includegraphics[width=0.5\linewidth, height=0.5\textheight]{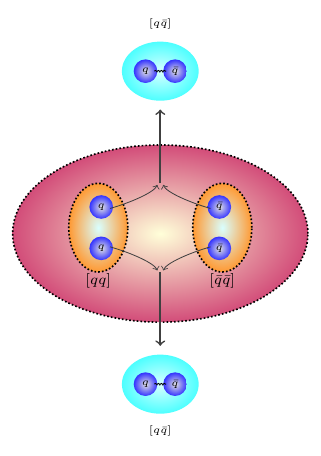}
	
	\caption[]{Pictorial representation of tetraquark $T_{qq\bar{q}\bar{q}}$ re-arrangement }
	
\end{figure}

The rearrangement of the $qq\bar{q}\bar{q}$ tetraquark yields two mesons, a process driven by the decay of the quark-antiquark pair ($q\bar{q}$) into lower mass states. This phenomenon elucidates the mechanism behind rearrangement decay. These decays manifest across various channels, including but not limited to:

1. The conversion and confinement of two gluons into light hadrons is obtained by the decay of color singlet spin-0 $q\bar{q}$ pair with a rate of order of $\alpha_{s}^{2}$. Similarly, the $q\bar{q}$ pair can also be converted into two photons. With the spectator $q\bar{q}$ pair in account, the decays observed are:
\begin{center}
	$qq\bar{q}\bar{q}\rightarrow\pi+\text{2 gluons} $ \\
	$qq\bar{q}\bar{q}\rightarrow\pi+2\gamma $
\end{center}

2. The conversion and confinement of three gluons into light hadrons is obtained by the decay of color singlet spin-1 pairs with a rate of order of $\alpha_{s}^{3}$. This color singlet spin- pair can also decay into three photons or two leptons. With the spectator $\rho$ meson in account, the decays observed are:
\begin{center}
	$qq\bar{q}\bar{q}\rightarrow\rho+\text{3 gluons}$ \\
	$qq\bar{q}\bar{q}\rightarrow\rho+\text{2 leptons}$ \\
	$qq\bar{q}\bar{q}\rightarrow\rho+3 \gamma$ \\
\end{center}

3. The spin-0 and spin-1 $q\bar{q}$ pair in the color octet representation annihilates into one gluon and two gluons, respectively. The decay observed is:
\begin{center}
	$qq\bar{q}\bar{q}\rightarrow\pi+\text{1 gluon} $ \\
	$qq\bar{q}\bar{q}\rightarrow \rho + \text{2 gluons} $
\end{center}

The ratio of overlap probabilities of the annihilating $q\bar{q}$ pair in $qq\bar{q}\bar{q}$ is proportional to the decay rates:
\begin{equation}
	\varrho = \frac{|\Psi_{qq\bar{q}\bar{q}}(0)|^{2}}{|\Psi_{\rho}(0)|^{2}}
\end{equation}

The individual decay rate is obtained using the simple formula \cite{Berestetskii:1982qgu}:
\begin{equation}
	\Gamma(i)^{spin}_{color}=||\Psi_{qq\bar{q}\bar{q}/sq\bar{q}\bar{q}}(0)|^{2}v\sigma(i)^{spin}_{color}\rightarrow f)
\end{equation}
where $|\Psi_{{qq\bar{q}\bar{q}}}(0)|^{2},v,$ and $\sigma$ are the overlap probability of the annihilating or decaying pair, relative velocity, and the spin-averaged annhilation/decay cross section in the final state $f$, respectively. The sum total of all the individual decay rates yields the total decay rate. The spectator pair $q\bar{q}$ can appear as $\pi$ or $\rho$ on the mass shell. 

The singlet spin 0 decay rate is given by:

\begin{equation}
	\begin{split}
		\Gamma_{1}&=\Gamma(qq\bar{q}\bar{q}\rightarrow\pi+\text{2 gluons}	)\\
		&=2\cdot\frac{1}{4}\cdot|\Psi_{qq\bar{q}\bar{q}}(0)|^{2} v \sigma((q\bar{q}^{0}_{1})\rightarrow \text{2 gluons}) \\
		& = \frac{1}{2}\Gamma(q\bar{q}^{0}_{1}\rightarrow \text{2 gluons})\cdot\varrho \\
	\end{split}
\end{equation}

\begin{equation}
	\begin{split}
		\Gamma_{2}&=\Gamma(qq\bar{q}\bar{q}\rightarrow\pi+ 2\gamma)\\
		&=2\cdot\frac{1}{4}\cdot|\Psi_{qq\bar{q}\bar{q}}(0)|^{2} v \sigma((q\bar{q}^{0}_{1})\rightarrow 2 \gamma) \\
		& = \frac{1}{2}\Gamma((q\bar{q}^{0}_{1})\rightarrow2\gamma)\cdot\varrho\\
	\end{split}
\end{equation}

Similarly, the color singlet spin 1 decay rate is given by

\begin{equation}
	\begin{split}
		\Gamma_{3}&=\Gamma(qq\bar{q}\bar{q} \rightarrow \rho + \text{3 gluons})\\
		&=2\cdot\frac{1}{12}\cdot|\Psi_{qq\bar{q}\bar{q}}(0)|^{2} v \sigma((q\bar{q}^{1}_{1})\rightarrow \text{3 gluons}) \\
		& = \frac{1}{6}\Gamma(q\bar{q}^{1}_{1}\rightarrow \text{3 gluons})\cdot\varrho \\
	\end{split}
\end{equation}

\begin{equation}
	\begin{split}
		\Gamma_{4}&=\Gamma(qq\bar{q}\bar{q} \rightarrow \rho + \text{3 photons})\\
		&=2\cdot\frac{1}{12}\cdot|\Psi_{qq\bar{q}\bar{q}}(0)|^{2} v \sigma((q\bar{q}^{1}_{1})\rightarrow \text{3 photons}) \\
		& = \frac{1}{6}\Gamma((q\bar{q}^{1}_{1})\rightarrow \text{3 photons})\cdot\varrho \\
	\end{split}
\end{equation}

\begin{equation}
	\begin{split}
		\Gamma_{5}&=\Gamma(qq\bar{q}\bar{q} \rightarrow \rho + \text{2 leptons})\\
		&=2\cdot\frac{1}{12}\cdot|\Psi_{qq\bar{q}\bar{q}}(0)|^{2} v \sigma((q\bar{q}^{1}_{1})\rightarrow \text{2 leptons}) \\
		& = \frac{1}{6}\Gamma((q\bar{q}^{1}_{1})\rightarrow \text{2 leptons})\cdot\varrho \\
	\end{split}
\end{equation}

The annihilation of the spin-1 octet $q\bar{q}$ pair into the light quark pair is given by:

\begin{equation}
	\begin{split}
		\Gamma_{6}&=\Gamma(qq\bar{q}\bar{q} \rightarrow \rho + \text{2 gluon})\\
		&= 2 \cdot\frac{1}{6} \cdot\frac{1}{4}\biggr( \frac{4\pi\alpha_{s}^{2}}{3}\frac{4}{m_{q\bar{q}}^{2}} \biggr)|\Psi_{q\bar{q}}(0)|^{2} \cdot\varrho\\
	\end{split}
\end{equation}
The annihilation of the spin-0 octet $s\bar{s}$ pair into the light quark pair is given by:
\begin{equation}
	\begin{split}
		\Gamma_{7}&=\Gamma(qq\bar{q}\bar{q} \rightarrow \pi + \text{1 gluon})\\
		&= 2 \cdot\frac{1}{6} \cdot\frac{1}{4}\biggr( \frac{4\pi\alpha_{s}^{2}}{m_{q\bar{q}}^{2}}\cdot18 \biggr)|\Psi_{q\bar{q}}(0)|^{2} \cdot\varrho\\
	\end{split}
\end{equation}

The leptonic decay of color singlet spin-0 $u\bar{d}$ pair is given by : 

\begin{equation}
	\begin{split}
		\Gamma_{8}&=\Gamma(qq\bar{q}\bar{q} \rightarrow \pi^{\pm} + e^{\mp}\nu_{e})\\
		&=2\cdot\frac{1}{4}\cdot|\Psi_{qq\bar{q}\bar{q}}(0)|^{2} v \sigma(\pi^{\pm}\rightarrow e^{\mp}\nu_{e}) \\
		&=\frac{1}{2} \frac{G_{F}^{2}}{8m_{\pi^{\pm}}^{3}\pi}f^{2}_{\pi^{\pm}}[m_{e}(m_{\pi^{\pm}}^{2}-m_{e}^{2})]^{2} \\
	\end{split}
\end{equation}

\begin{equation}
	\begin{split}
		\Gamma_{9}&=\Gamma(qq\bar{q}\bar{q} \rightarrow \pi^{\pm} + \mu^{\mp}\nu_{\mu})\\
		&=2\cdot\frac{1}{4}\cdot|\Psi_{qq\bar{q}\bar{q}}(0)|^{2} v \sigma(\pi^{\pm}\rightarrow \mu^{\mp}\nu_{\mu}) \\
		&=\frac{1}{2} \frac{G_{F}^{2}}{8m_{\pi^{\pm}}^{3}\pi}f^{2}_{\pi^{\pm}}[m_{\mu}(m_{\pi^{\pm}}^{2}-m_{\mu}^{2})]^{2} \\
	\end{split}
\end{equation}

The rare leptonic decay of color singlet spin-1 $u\bar{d}$ pair is given by : 

\begin{equation}
	\begin{split}
		\Gamma_{10}&=\Gamma(qq\bar{q}\bar{q} \rightarrow \rho^{\pm} + \gamma e^{\mp}\nu_{e})\\
		&=2\cdot\frac{1}{12}\cdot|\Psi_{qq\bar{q}\bar{q}}(0)|^{2} v \sigma((\rho^{\pm})\rightarrow \gamma e^{\mp}\nu_{e}) \\
		&=\frac{1}{6}\frac{\alpha G_{F}^{2}}{2592\pi^{2}}|V_{du}|^{2}f^{2}_{\pi^{\pm}}m_{\pi^{\pm}}^{3}(x_{u}+x_{d}) \\
	\end{split}
\end{equation}

\begin{equation}
	\begin{split}
		\Gamma_{11}&=\Gamma(qq\bar{q}\bar{q} \rightarrow \rho^{\pm} + \gamma \mu^{\mp}\nu_{\mu})\\
		&=2\cdot\frac{1}{48}\cdot|\Psi_{qq\bar{q}\bar{q}}(0)|^{2} v \sigma(\rho^{\pm}\rightarrow \gamma \mu^{\mp}\nu_{\mu}) \\
		&=\frac{1}{24}\frac{\alpha G_{F}^{2}}{2592\pi^{2}}|V_{du}|^{2}f^{2}_{\pi^{\pm}}m_{\pi^{\pm}}^{3}(x_{u}+x_{d})\\
	\end{split}
\end{equation}
where ,
\begin{equation}
	x_{u} = \biggr(3-\frac{M_{\pi^{\pm}}}{m_{u}} \biggr)^{2}, \;\;\;\; x_{s} = \biggr(3-2\frac{M_{\pi^{\pm}}}{m_{d}} \biggr)^{2}.
\end{equation}

where $G_{F}$ is the Fermi constant, $m_{u}$ is the mass of the up quark, $M_{K^{\pm}}$ is the mass of kaon, $f_{K^{\pm}}$ is the weak decay constant, and $|V_{su}|$ is the element of the CKM matrix, respectively. The branching ratio of various decay channels of $T_{qq\bar{q}\bar{q}}$ and $T_{sq\bar{q}\bar{q}}$ tetraquark in antitriplet-triplet configuration have been tabulated in table \ref{spectatordecay}. 

\begin{table*}
	\centering
	\caption{Decay widths for various decay channels in spectator model for $T_{^{1}S_{0}}$}
	\label{spectatordecay}
	\begin{tabular}{ccc}
		\hline
		\multirow{2}{*}{State}  &\multicolumn{2}{c}{$\bar{\textbf{3}}-\textbf{3}$}  \\
		& Semi-Relativistic & Non-Relativistic \\
		\hline
		$qq\bar{q}\bar{q}\rightarrow\pi+\text{ 2 gluons}$ &$1.292\times10^{6}$&$1.249\times10^{6}$\\
		$qq\bar{q}\bar{q}\rightarrow\pi+ \text{2 photons}$ &$2.192\times10^{6}$ &$1.943\times10^{6}$\\
		$qq\bar{q}\bar{q}\rightarrow\pi^{\mp}+e^{\pm}+\nu_{e}$ &$149.161$&$202.576$\\
		$qq\bar{q}\bar{q}\rightarrow\pi^{\mp}+\mu^{\pm}+\nu_{\mu}$  &$121.077\times10^{3}$&$164.434\times10^{3}$\\
		$qq\bar{q}\bar{q}\rightarrow\pi+\text{ 1 gluon}$ &$36.471\times10^{3}$&$32.386\times10^{3}$\\
		$qq\bar{q}\bar{q}\rightarrow\rho+\text{ 3 gluons}$ &$130.851\times10^{6}$&$116.846\times10^{6}$\\
		$qq\bar{q}\bar{q}\rightarrow\rho+\text{2 leptons}$ &$281.902\times10^{3}$&$382.854\times10^{3}$\\
		$qq\bar{q}\bar{q}\rightarrow\rho+\text{3 photons}$ &$0.644\times10^{6}$&$0.555\times10^{6}$\\
		$qq\bar{q}\bar{q}\rightarrow \rho^{\pm}+\gamma e^{\mp}\nu_{e}$&$92.816$&$126.053$\\
		$qq\bar{q}\bar{q}\rightarrow \rho^{\pm}+\gamma\mu^{\mp}\nu_{\mu}$ &$92.816$&$126.053$\\
		$qq\bar{q}\bar{q}\rightarrow\rho+\text{ 2 gluons}$ &$2.695\times10^{3}$&$2.393\times10^{3}$\\
		
		\hline
	\end{tabular}
\end{table*}

\section{Regge trajectories}

\begin{figure*}[t]
	\centering
	\begin{subfigure}{0.45\textwidth}
		\includegraphics[width=1.05\linewidth, height=0.3\textheight]{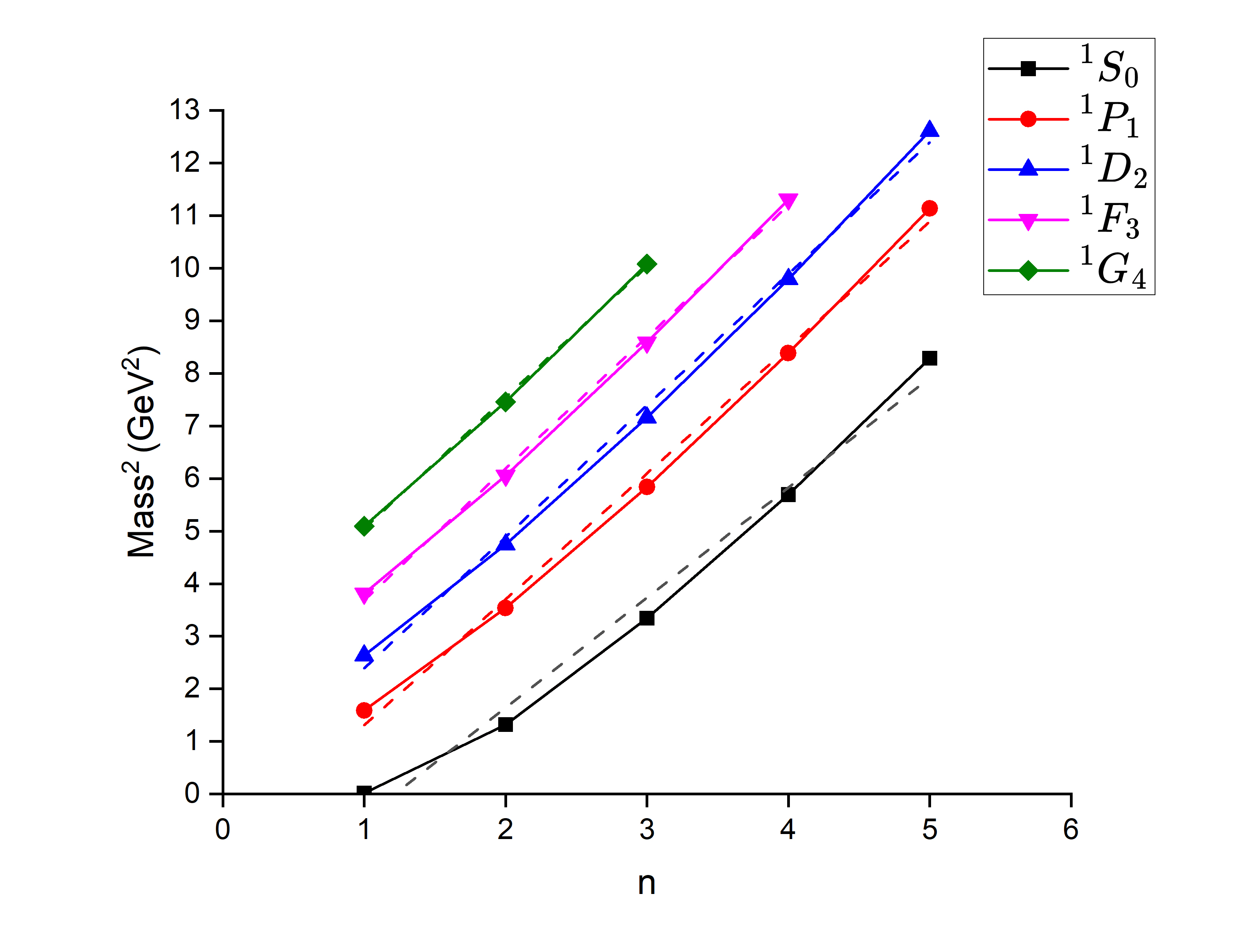}
		\caption{Non-Relativistic formalism}
		\label{fig:mesonnrgraph1nat}
	\end{subfigure}
	\begin{subfigure}{0.45\textwidth}
		\includegraphics[width=1.05\linewidth, height=0.3\textheight]{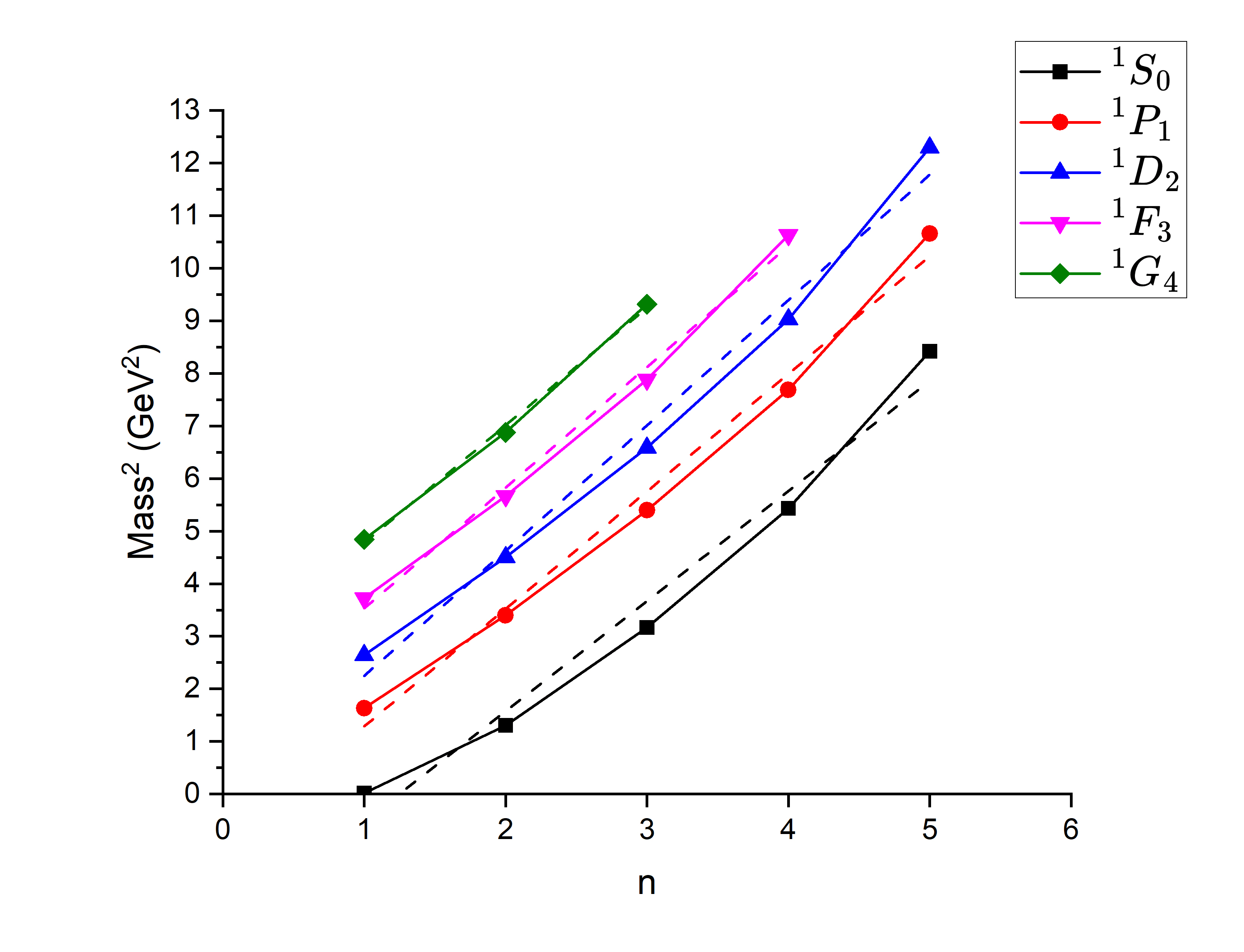}
		\caption{Semi-Relativistic formalism}
		\label{fig:mesonsrgraph1nat}
	\end{subfigure}
	\caption[]{Regge trajectory in the $(n, M^{2})$ plane for $\pi$ meson with unnatural parity, (Spin S = 0)}
	\end{figure*}

\begin{figure*}[t]
	\centering
	\begin{subfigure}{0.475\textwidth}
		\includegraphics[width=1\linewidth, height=0.3\textheight]{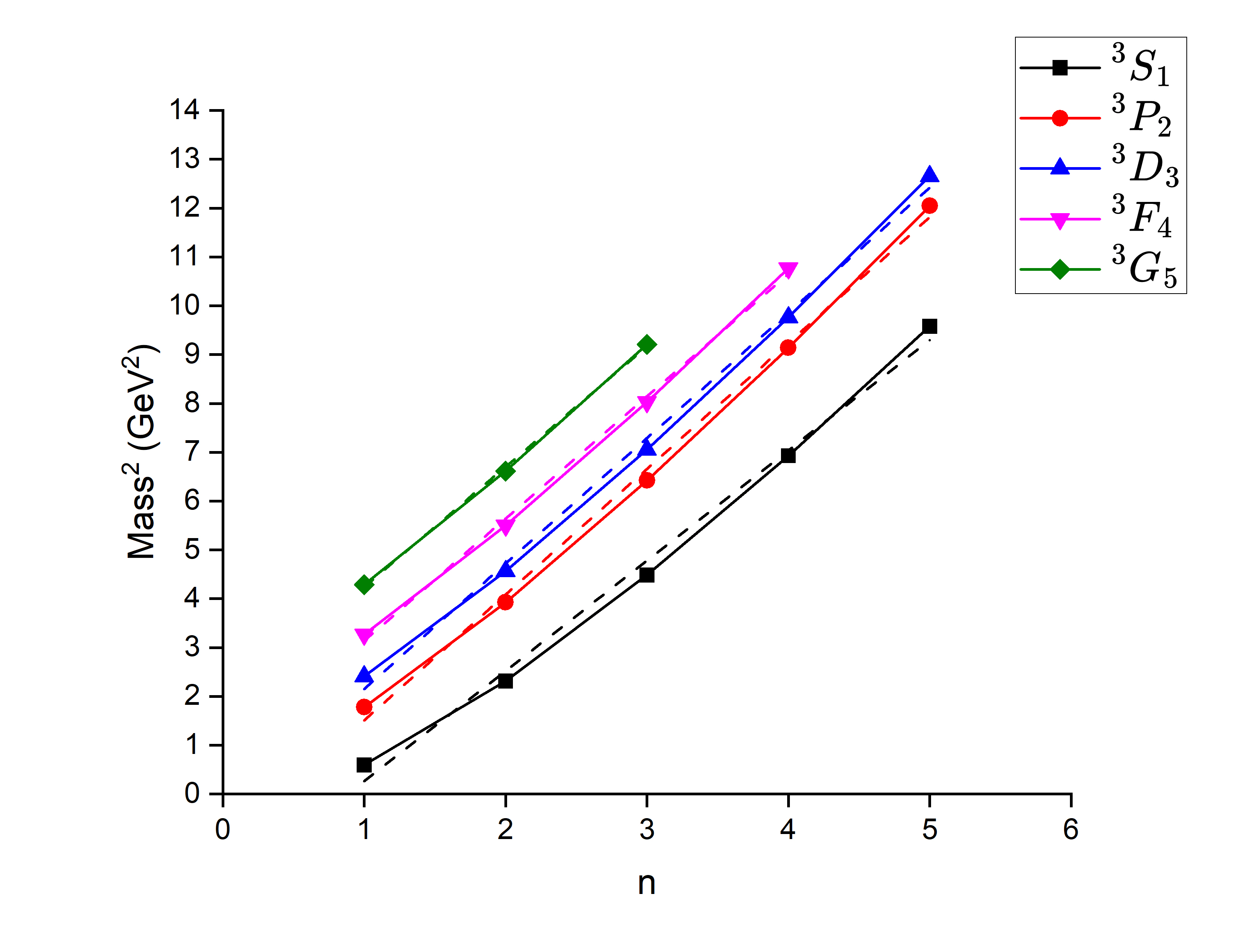}
		\caption{Non-Relativistic formalism}
		\label{fig:mesonnrgraph2nat}
	\end{subfigure}
	\begin{subfigure}{0.475\textwidth}
		\includegraphics[width=1\linewidth, height=0.3\textheight]{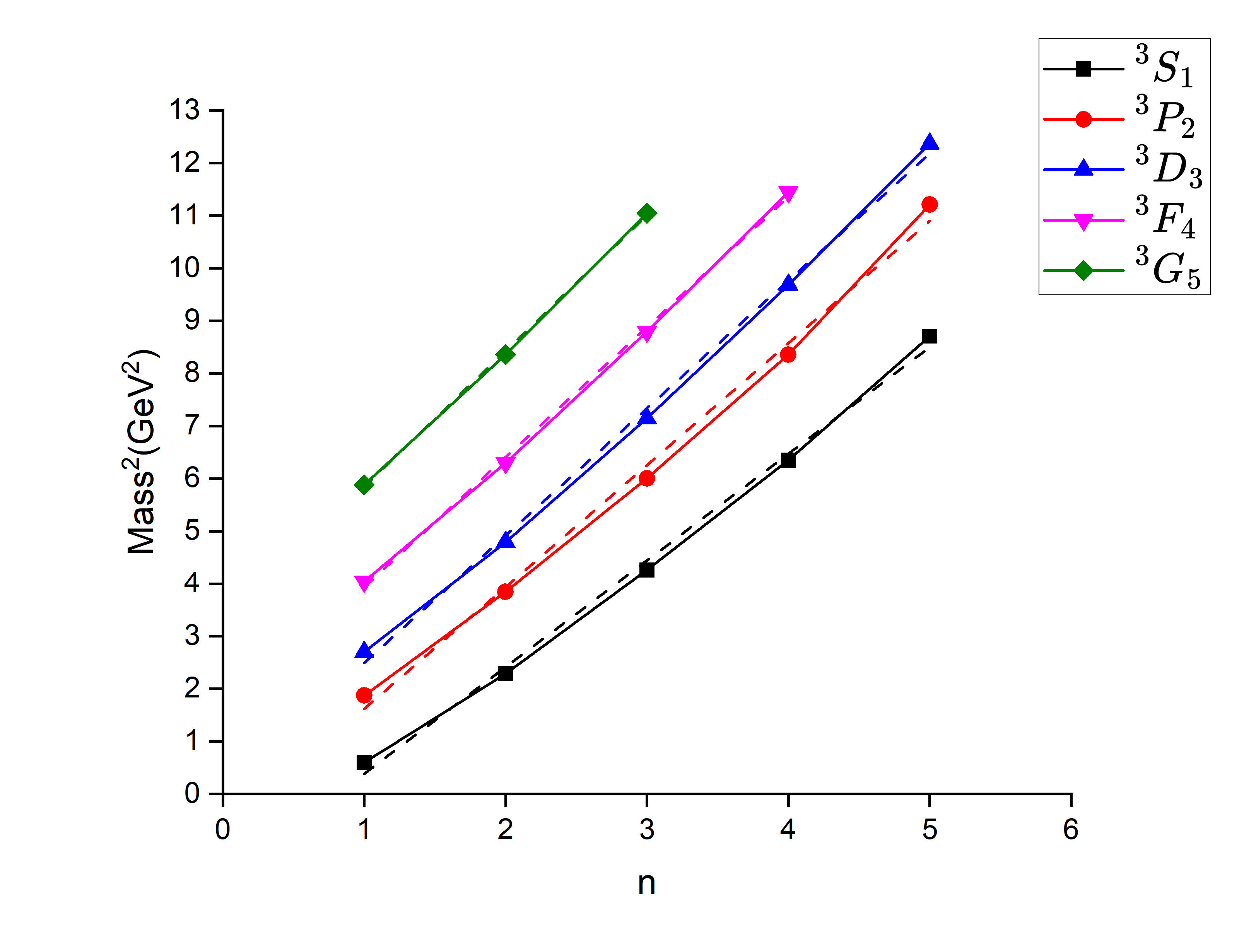}
		\caption{Semi-Relativistic formalism}
		\label{fig:mesonsrgraph2nat}
	\end{subfigure}
	\caption[]{Regge trajectory in the $(n, M^{2})$ plane for $\pi$ meson with natural parity, (Spin S = 1)}
\end{figure*}

\begin{figure*}[t]
	\centering
	\begin{subfigure}{0.475\textwidth}
		\includegraphics[width=1\linewidth, height=0.3\textheight]{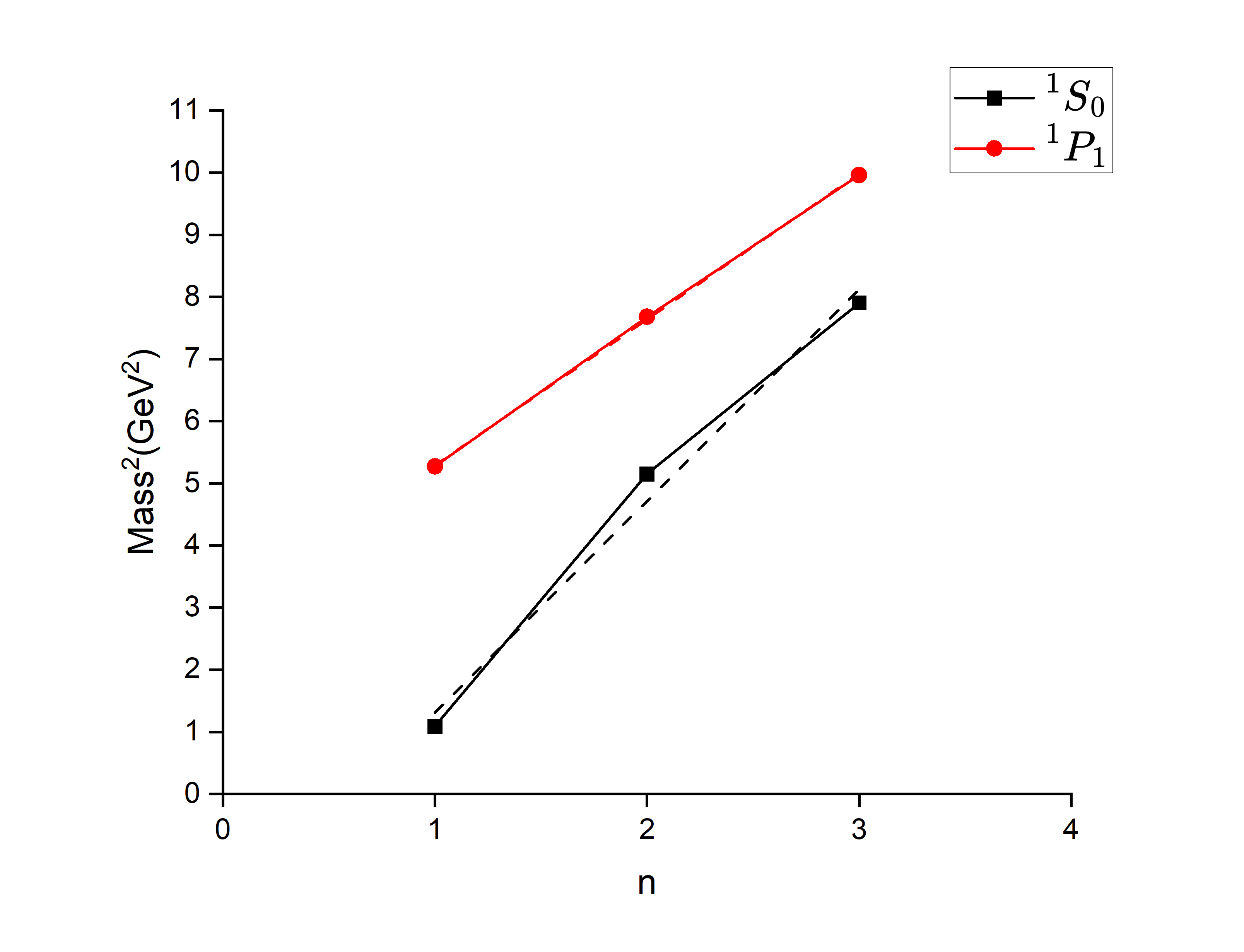}
		\caption{Spin S = 0}
		\label{fig:tetraS0}
	\end{subfigure}
	\begin{subfigure}{0.475\textwidth}
		\includegraphics[width=1\linewidth, height=0.3\textheight]{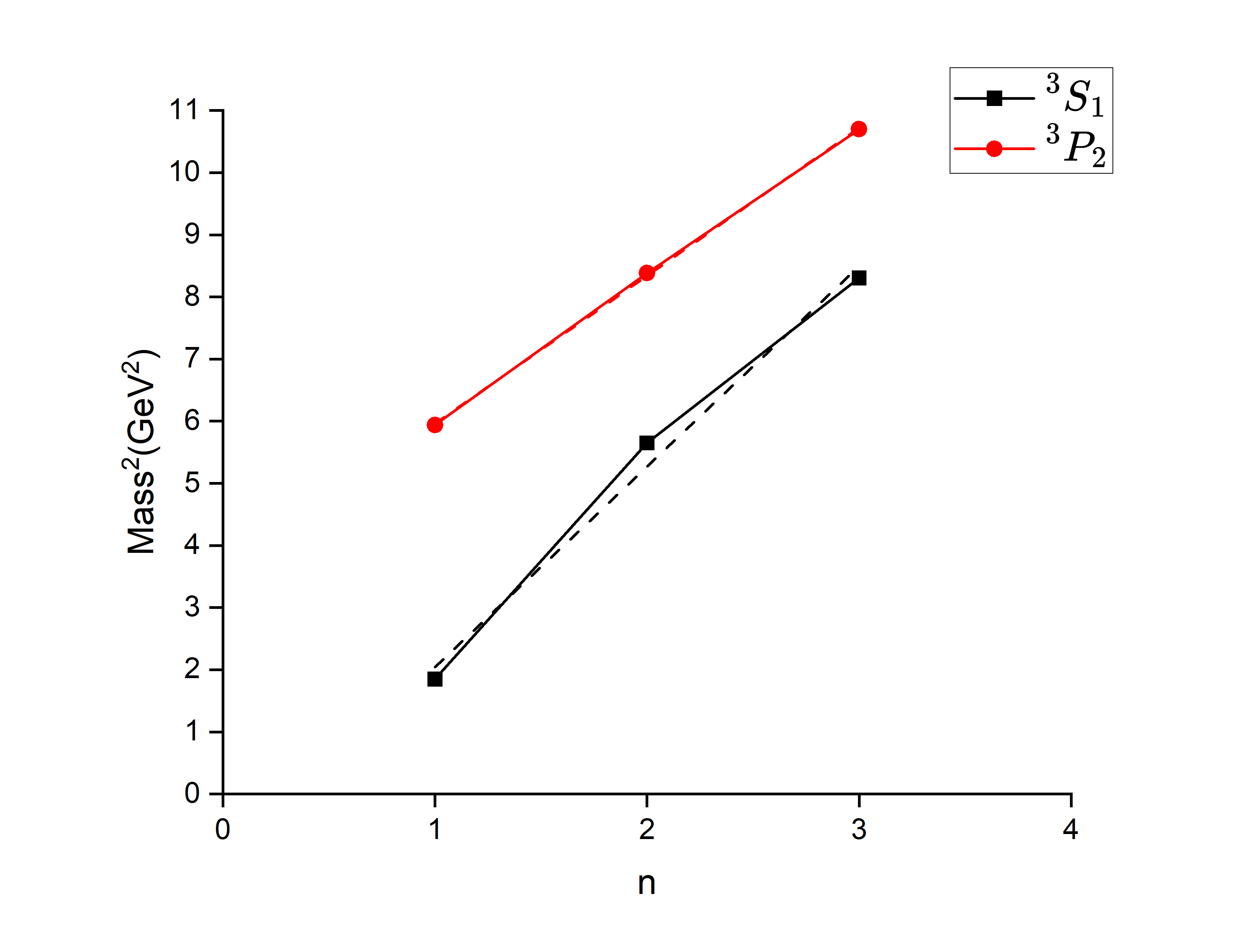}
		\caption{Spin S = 1}
		\label{fig:tetraS1}
	\end{subfigure}
		\begin{subfigure}{0.475\textwidth}
		\includegraphics[width=1\linewidth, height=0.3\textheight]{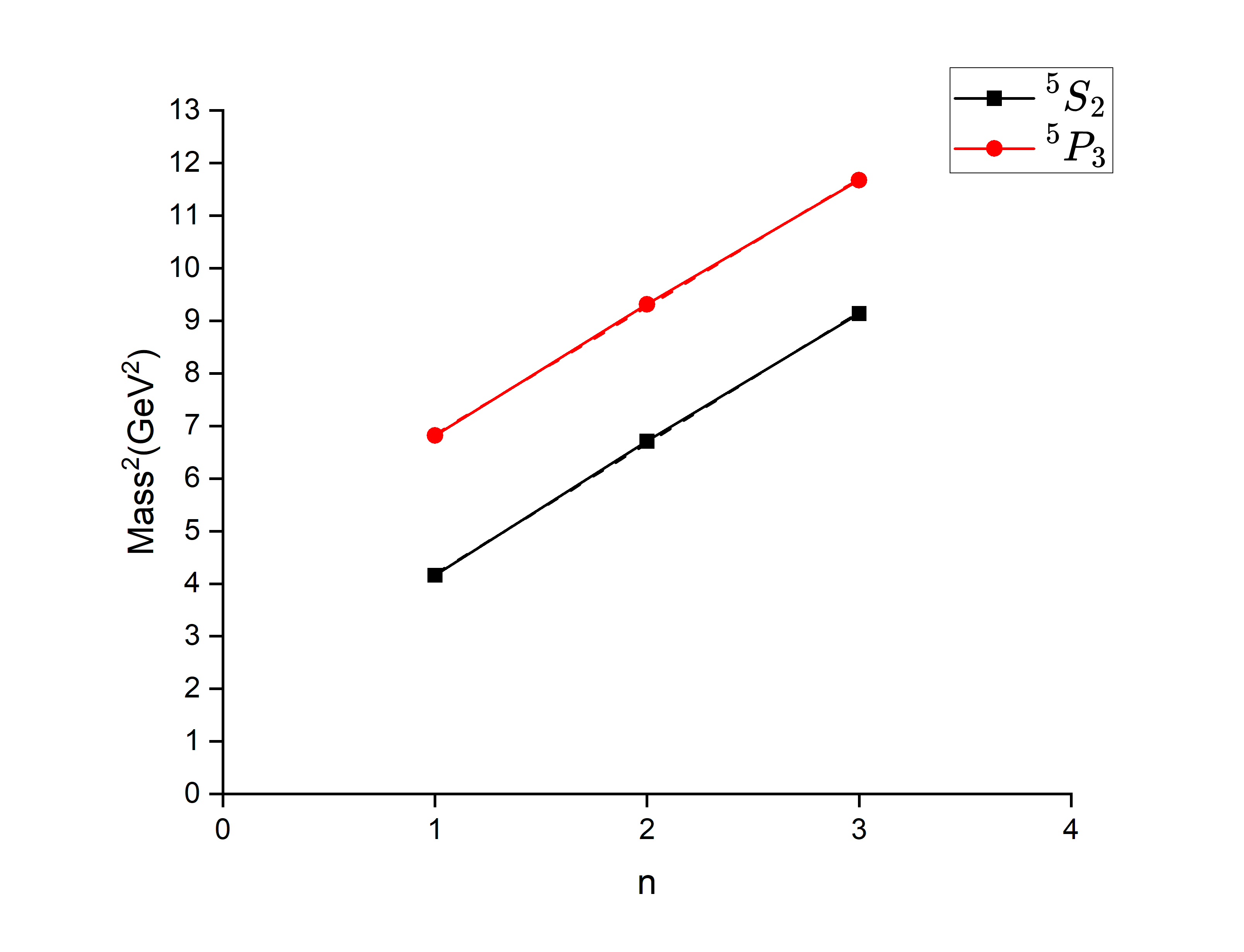}
		\caption{Spin S = 2}
		\label{fig:tetraS2}
	\end{subfigure}
	\caption[]{Regge trajectory in the $(n, M^{2})$ plane for tetraquarks with Spins S = 0,1 and 2}
\end{figure*}

This section discusses regge trajectories of the calculated mass spectrum of $\pi$ mesons and all light tetraquarks. The Regge trajectories in the $(n,M^{2})$ plane are plots of the principal quantum number, n, against the square of the resonance mass, $M^{2}$. The Regge trajectories in the $(n, M^{2})$ plane are drawn for the $\pi$ meson with natural and unnatural parity states, as shown in Figs. \ref{fig:mesonnrgraph1nat}, \ref{fig:mesonsrgraph1nat}, \ref{fig:mesonnrgraph2nat} and \ref{fig:mesonsrgraph2nat}. The solid straight lines represent our calculated results. Similarly, the trajectories in the $(n,M^{2})$ plane are depicted in figs. \ref{fig:tetraS0}, \ref{fig:tetraS1}, and \ref{fig:tetraS2} for all light tetraquarks with various spins.

\section{Results and Discussions}

In this work, the mass spectra of $\pi$ mesons are calculated and presented in Tables \ref{SWaveMesonmass} and \ref{GWaveMesonmass}. Using the fitting parameters derived from the $\pi$ meson mass spectrum, the masses of diquarks and anti-diquarks are computed for various color configurations. The mass spectroscopy of all strange tetraquarks is then determined within both semi-relativistic and non-relativistic frameworks using the diquark-antidiquark formalism, considering various possible internal structures, which are shown in Tables \ref{Swavetriplet}. Table \ref{twomesonthreshold} displays the two-meson thresholds for various tetraquark states. Additionally, the decay properties of tetraquarks in different decay channels are calculated. Various decay channels for $T_{qq\bar{q}\bar{q}}$ tetraquarks are explored using Fierz rearrangement, the spectator model, and heavy quarkonium annihilation, with the results presented in Tables \ref{annihilationdecay} and \ref{spectatordecay}. 

\subsection{Meson}

The calculated pionic mass spectra, when compared with various theoretical studies and with the PDG data, show good agreement. Numerous resonances that match the description of calculated states are discussed here.

\subsubsection*{\textbf{A. $^{1}S_{0}$ State}}

The mass description of the five resonances listed in PDG matches the mass spectra for the $^{1}S_{0}$ pion state. The $\pi^{\pm}$ is a well-established state that was first observed in 1947 
\cite{Lattes:1947mw}. Since its first observation, the ground state $\pi$ meson has been observed in numerous processes with an average mass of $139.57039\pm0.00018$ MeV and a mean life of $2.6033\pm0.0005\times10^{-8}s$. This state has an $I^{G}(J^{P})$ value of $1^{-}(0^{-})$. The present study estimates the mass of this state at 139.60 MeV and 139.49 MeV for non-relativistic and semi-relativistic formalism, respectively. The compared theoretical studies also predict similar results. Study \cite{Godfrey:1985xj}, \cite{Ebert:2009ub} and \cite{Kojo:2022psi} predict it a few MeVs higher than PDG, while study \cite{Ishida:1986vn}, \cite{Vijande:2004he} and \cite{Fischer:2014xha} predict it within 1 MeV mass difference with that of PDG. 

The $\pi(1800)$ resonance has been observed in $190\pi^{-}p\rightarrow\pi^{-}\pi^{+}\pi^{-}p$ \cite{COMPASS:2018uzl}, $18\pi^{-}p\rightarrow\eta\eta\pi^{-}p$ \cite{E852:2008eut}, $18.3\pi^{-}p\rightarrow\pi^{+}\pi^{-}\pi^{-}p$ \cite{Chung:2002pu},  $37\pi^{-}A\rightarrow\eta\eta\pi^{-}A$ \cite{Amelin:1995fg}, $36\pi^{-}A\rightarrow\pi^{+}\pi^{-}\pi^{-}A$ \cite{VES:1995mdc} and many more such processes with an average mass of $1810^{+9}_{-11}$ MeV and an average decay width of $215^{+7}_{-8}$ MeV. Similarly, the $X(1835)$ resonance has been observed in $J/\psi\rightarrow\gamma\pi^{+}\pi^{-}\eta^{'}$ \cite{BESIII:2016fbr}, $J/\psi\rightarrow\gamma K^{0}_{s}K^{0}_{s}\eta$ \cite{BESIII:2015xco} and many other processes with an average mass of $ 1826.5^{+13}_{-3.4}$ MeV and an average decay width of $242^{+14}_{-15}$ MeV. Both of these resonances have $J^{PC}$ values of $0^{-+}$, which match with the $^{1}S_{0}$ state; however, the $I^{G}$ value of $\pi(1800)$ is $1^{-}$ while that of $X(1835)$ is undefined. The masses of both of these resonances fit very well with the mass of the $3 ^{1}S_{0}$ pion state of the present work. The compared studies \cite{Godfrey:1985xj} and \cite{Ishida:1986vn} predict the mass of the $3 ^{1}S_{0}$ state to be higher than mass in both formalisms. However, studies \cite{Ebert:2009ub} and \cite{Fischer:2014xha} match the mass description in semi-relativistic formalism, while study \cite{Kojo:2022psi} matches the mass in non-relativistic formalism. Ref \cite{Vijande:2004he}, \cite{Ebert:2009ub}, \cite{Kojo:2022psi} and \cite{Fischer:2014xha} also predict the $3 ^{1}S_{0}$ pion state to be $\pi(1800)$ meson, while ref. \cite{Ishida:1986vn} predicts it to be $\pi(1770)$.

The $\pi(2360)$ resonance has been observed through partial wave analysis of $p\bar{p}$ annihilation channels in flight, with quantum numbers $I = 1$ and $C = +1$ with an average mass of $2360 \pm 25$ MeV and a decay width of $300^{+100}_{-50}$ MeV \cite{Anisovich:2001pn}. This analysis involved the processes $2.0 p\bar{p} \rightarrow 3\pi^{0}, \pi^{0}\eta, \pi^{0}\eta'$. The $I^{G}(J^{PC})$ value for this resonance is $1^{-}0^{-+}$. The $X(2340)$ resonance has been observed in the $15\pi^{+}p\rightarrow p5\pi$ process with an average mass of $2340\pm20$ MeV and a decay width of $180\pm60$ MeV \cite{Baltay:1975wx}. The $I^{G}(J^{PC})$ value for this resonance is undefined. Both of these resonances fit the mass description of the $3 ^{1}S_{0}$ pion state for the present study in both formalisms. However, since $\pi(2360)$ has a better defined $I^{G}(J^{PC})$ value, it is a more prominent candidate for the state. All the compared studies predict mass lower for this state, with reference to \cite{Ebert:2009ub} and \cite{Kojo:2022psi}, which predict this state to be the $\pi(2070)$ resonance. 

\subsubsection*{\textbf{B. $^{3}S_{1}$ State}}

The vector state $\rho(770)$ is a well-established state with $I^{G}J^{PC}$ value $1^{+}(1^{--})$. This state has been detected through various processes, including $e^{+}e^{-}$ annihilation, $\tau$ decay, hadroproduction, and photoproduction, with corresponding average masses of $775.26 \pm 0.23$ MeV, $775.11 \pm 0.34$ MeV, $766.5 \pm 1.1$ MeV, $769.2 \pm 0.9$ MeV, and $769.0 \pm 0.9$ MeV \cite{ParticleDataGroup:2022pth}. The average decay widths for these processes are $147.4 \pm 0.8$ MeV, $149.1 \pm 0.8$ MeV, $150 \pm 2.4$ MeV, $151.5^{+1.9}_{-2.1}$ MeV, and $150.9 \pm 1.7$ MeV, respectively. The estimated mass for this state in both forms of the present study matches the PDG values. Ref. \cite{Godfrey:1985xj}, \cite{Vijande:2004he} and \cite{Ebert:2009ub} also predict values near the PDG data for this state. However, ref. \cite{Ishida:1986vn}, \cite{Kojo:2022psi} and \cite{Fischer:2014xha} predict it lower by a few MeVs. 

The $2 ^{3}S_{1}$ state has three potential candidates: $\rho(1450)$, $\rho(1570)$, and $X(1575)$. The $X(1575)$ resonance has been observed in the $J/\psi \rightarrow K^{+}K^{-}\pi^{0}$ process with an average mass of $1576^{+49+98}_{-55-91}$ MeV and a decay width of $818^{+22+64}_{-23-133}$ MeV \cite{BES:2006kmo}. The $\rho(1570)$ resonance has an average mass of $1570 \pm 70$ MeV and quantum numbers $I^{G}(J^{PC}) = 1^{+}(1^{--})$. This state has been observed in the $10.6 \, e^{+}e^{-} \rightarrow \phi \pi^{0} \gamma$ process with an average decay width of $144 \pm 90$ MeV \cite{BaBar:2007ceh}. The $\rho(1450)$ resonance, which also has $I^{G}(J^{PC}) = 1^{+}(1^{--})$, is observed in various processes, featuring an average mass of $1465 \pm 25$ MeV and an average decay width of $400 \pm 60$ MeV \cite{ParticleDataGroup:2022pth}. Except for ref. \cite{Ishida:1986vn}, all compared studies predict the mass for this state to be at least 50 MeV less than the present study, and most identify this state as the $\rho(1450)$ resonance. Reference \cite{Ishida:1986vn} alone predicts it to be the $\rho(1570)$ resonance. Therefore, it is reasonable to lean towards identifying it as the $\rho(1450)$. 
The radially excited states of S-wave pions, particularly the $4S$ and $5S$ states, predict masses higher than those reported in other studies.

\subsubsection*{\textbf{C. $^{1}P_{1}$ State}}

The $b_{1}(1235)$ state is a well-defined pionic resonance with an average mass of $1229.5\pm3.2$ MeV and a $I^{G}(J^{PC}) = 1^{+}(1^{+-})$. It has been observed in several processes, such as $\bar{p}p\rightarrow2\pi^{+}2\pi^{-}\pi^{0}$ \cite{ASTERIX:1993wam}, $38,100\pi^{-}p\rightarrow\omega\pi^{0}n$ \cite{IHEP-IISN-LANL-LAPP-KEK:1992puu}, $8.95 \pi^{-}p \omega\pi^{0}n$ \cite{Fukui:1990ki}, $25-55\gamma p\rightarrow\omega\pi X$ \cite{OmegaPhoton:1984ols}, $12\pi^{-}p\rightarrow \omega\pi p$ \cite{Evangelista:1980xe}, $11\pi^{-}p\rightarrow \pi^{-} \omega p$ \cite{Gessaroli:1977mt} and many more with an averaged decay width of $142\pm9$ MeV. The mass description of this resonance matches closely with the $1 ^{1}P_{1}$ state in the current work under both formalisms. Apart from reference \cite{Fischer:2014xha}, all the studies compared predict masses around 1250 MeV for this state and predict the resonance $b_{1}(1235)$ as the state, whereas reference \cite{Fischer:2014xha} predicts a significantly lower mass. 

\subsubsection*{\textbf{D. $^{3}P_{0}$ State}}

The $2 ^{3}P_{0}$ state of the present study matches the description of $a_{0}(1450)$ resonance, which has 
$I^{G}(J^{PC}) = 1^{-}(0^{++})$. This resonance is a well-established state that is observed in $0.0p\bar{p}\rightarrow K^{0}_{L}K^{\pm}\pi^{\mp}$ \cite{Abele:1998qd} and $23 \pi^{-}p\rightarrow n2K^{0}_{S}$ \cite{Etkin:1982se} with an average mass of $1439\pm34$ MeV and an average decay width of $258\pm14$ MeV. The current study predicts the mass lower than the compared studies, except ref. \cite{Fischer:2014xha}, which predicts it even lower. The compared studies do not predict any resonance for the given state.

The $a_{0}(2020)$ and $a_{0}(1950)$ resonances are both good candidates for the $3 ^{3}P_{0}$ state of pion in semi-relativistic formalism. The $a_{0}(1950)$ resonance has an average mass of $1931\pm26$ MeV and has been seen in the $\gamma\gamma\rightarrow\eta_{c}(1S)\rightarrow K\bar{K}\pi$ process with an average decay width of $271\pm40$ MeV \cite{BaBar:2015kii}. Similarly, the $a_{0}(2020)$ resonance has been seen in the $\bar{p}p\rightarrow\pi^{0}\eta^{'},\pi^{0}\eta$ process with an average mass of $2025\pm30$ MeV and decay width of $330\pm75$ MeV \cite{CrystalBarrel:1999zaz}. Both the resonances are not well established and have $I^{G}(J^{PC}) = 1^{-}(0^{++})$. Ref. \cite{Ishida:1986vn} predicts a higher mass for this state, while ref. \cite{Fischer:2014xha} predicts a lower mass. Ref \cite{Ebert:2009ub} predicts a mass similar to the semi-relativistic mass of the current study and predicts it to be the $a_{0}(2020)$ resonance. 

\subsubsection*{\textbf{E. $^{3}P_{1}$ State}}

The $a_{1}(1260)$ resonance is a well-established particle with quantum numbers $I^{G}(J^{PC}) = 1^{-}(1^{++})$. This resonance has been observed in the $190\pi^{-}p \rightarrow \pi^{-}\pi^{+}\pi^{-}p$ process with an average mass of $1230 \pm 40$ MeV and an average decay width of $420 \pm 35$ MeV \cite{COMPASS:2018uzl}. This description aligns well with the $1 ^{3}P_{1}$ state of the pion in the current study using non-relativistic formalism. Almost all compared studies, except for reference \cite{Fischer:2014xha}, estimate the mass of this state to be nearly 50 MeV less than the present work and identify it as the $a_{1}(1260)$ resonance.

The $a_{1}(1930)$ resonance is not a firmly established particle and has quantum numbers $I^{G}(J^{PC}) = 1^{-}(1^{++})$. This resonance has been detected in the partial wave analysis of the $p\bar{p}$ annihilation channel in flight with $I=1$ and $C=+1$, specifically in the $2.0\bar{p}p\rightarrow3\pi^{0},\pi^{0}\eta,\pi^{0}\eta^{'}$ process \cite{Anisovich:2001pn}. It has an average mass of $1930^{+30}_{-70}$ MeV and an average decay width of $155\pm45$ MeV. Another resonance, $a_{1}(2095)$ has similar quantum numbers and is seen in $18\pi^{-}p\rightarrow\eta\pi^{+}\pi^{-}\pi^{-}p$ with a mass of $2096\pm17\pm121$ MeV and decay width of $451\pm41\pm81$ MeV. These two resonances are ideal candidates for the $2 ^{3}P_{1}$ state of pion in the present study. The mass prediction of the compared studies for this state is diverse. Ref. \cite{Godfrey:1985xj} predicts it 100 MeV lower than the present study; ref. \cite{Ebert:2009ub} predicts it 150 MeV lower; and ref. \cite{Vijande:2004he} and \cite{Ebert:2009ub} predict it 250 MeV lower. On the other hand, ref. \cite{Ishida:1986vn} predicts it within 10 MeV of the present study. Ref. \cite{Vijande:2004he}, \cite{Ebert:2009ub} and \cite{Kojo:2022psi} assigns $a_{1}(1640)$ resonance to this state. 

\subsubsection*{\textbf{F. $^{3}P_{2}$ State}}
The $a_{2}(1320)$ resonance is a well-established particle with quantum numbers $I^{G}(J^{PC}) = 1^{-}(2^{++})$ and an average mass of $1318.2\pm0.6$ MeV. This resonance exhibits diverse decay widths, with its primary decay channel being $3\pi$, having an average decay width of $105.0^{+1.7}_{-1.9}$ MeV \cite{ParticleDataGroup:2022pth}. The $a_{2}(1320)$ resonance is considered a strong candidate for the ground state $^{3}P_{2}$ state of the pion, according to the present study. All compared studies predict the mass of the $1 ^{3}P_{2}$ state of the pion to be near 1320 MeV and assign this resonance for that state.

The $2 ^{3}P_{2}$ pion state has three resonances as potential candidates under the current study, namely $a_{2}(1950)$, $a_{2}(1990)$ and $a_{2}(2030)$. All three resonances have $I^{G}(J^{PC}) $ value$ 1^{-}(2^{++})$. The $a_{2}(1950)$ and $a_{2}(2030)$ resonances have been observed in the partial wave analysis of the $p\bar{p}$ annihilation channel in flight with $I=1$ and $C=+1$, specifically in the $1.96-2.41\bar{p}p$ process \cite{Anisovich:2001pn}. The $a_{2}(1950)$ resonance has an average mass of $1950^{+30}_{-70}$ MeV and a decay width of $180^{+30}_{-70}$ MeV. Similarly, the $a_{2}(2030)$ resonance has an average mass of $2030\pm20$ MeV and a decay width of $205\pm30$ MeV. The $a_{2}(1990)$ resonance was seen in $\gamma\gamma\rightarrow\pi^{+}\pi^{-}\pi^{0}$ \cite{Shchegelsky:2006es} and $18\pi^{-}p\rightarrow\omega\pi^{-}\pi^{0}p$ \cite{E852:2004rfa} processes with a mass of $2003\pm10\pm19$ MeV and a decay width of $249\pm23\pm32$ MeV. The compared studies except ref \cite{Ishida:1986vn} predict the mass of $2 ^{3}P_{2}$ state nearly 150 MeV less than that of the present study and assign it to be $a_{2}(1700)$ resonance. Ref. \cite{Ishida:1986vn} predicts the mass near the present study. Following the trend of S wave pions, radially excited states of P wave pions also predict masses higher than those reported in other studies.

\subsubsection*{\textbf{G. $^{1}D_{2}$ State}}

The $\pi_{2}(1670)$ resonance is a well-established particle with quantum numbers $I^{G}(J^{PC}) = 1^{-}(2^{-+})$ and an average mass of $1670.6^{+2.9}_{-1.2}$ MeV and an average decay width of $258^{+8}_{-9}$ MeV. This resonance has been observed in $190\pi^{-}p\rightarrow\pi^{-}\pi^{+}\pi^{-}p$ \cite{COMPASS:2018uzl}, $18\pi^{-}p\rightarrow\omega\pi^{-}\pi^{0}p$ \cite{E852:2004rfa}, $18.3\pi^{-}p\rightarrow\pi^{+}\pi^{-}\pi^{-}p$ \cite{Chung:2002pu}, $450pp\rightarrow p_{f}3\pi^{0}p_{s}$ \cite{WA102:2001rbg} and many more processes. This resonance is an apt fit for the $1 ^{1}D_{2}$ pion state of the current study. While ref. \cite{Vijande:2004he} and \cite{Kojo:2022psi} estimate the mass for this state a few MeVs less than the present study, ref. \cite{Godfrey:1985xj}, \cite{Ishida:1986vn} and \cite{Ebert:2009ub} estimate it a few MeVs higher. However, all the studies predict the state to be the $\pi_{2}(1670)$ resonance. 

The $\pi_{2}(2100)$ resonance also has $I^{G}(J^{PC}) = 1^{-}(2^{-+})$, but is not a well-established particle. This resonance has been observed in $36\pi^{-}A\rightarrow\pi^{+}\pi^{-}\pi^{-}A$ \cite{VES:1995mdc} and $63,94\pi p\rightarrow3\pi X$ process \cite{ACCMOR:1980llh} with an average mass of $2090\pm29$ MeV and an average decay width of $625\pm50$ MeV. This description fits the $2 ^{1}D_{2}$ state in the semi-relativistic formalism of the present study very well. Ref. \cite{Godfrey:1985xj} shows a similar mass to the non-relativistic formalism, while ref. \cite{Ishida:1986vn} estimate it 100 MeV higher and ref. \cite{Ebert:2009ub} and \cite{Kojo:2022psi} estimate it 50 MeV lower. Ref. \cite{Ishida:1986vn} predicts this state to be the $\pi_{2}(2100)$ resonance, while ref. \cite{Ebert:2009ub} and \cite{Kojo:2022psi} predict it to be the $\pi_{2}(2005)$ resonance.

\subsubsection*{\textbf{H. $^{3}D_{1}$ State}}

The $\rho(1570)$ resonance is not a well-established particle with quantum numbers $I^{G}(J^{PC}) = 1^{+}(1^{--})$. This resonance has been observed in the $10.6e^{+}e^{-} \rightarrow \phi\pi^{0}\gamma$ process with an average mass of $1570 \pm 70$ MeV and an average decay width of $144 \pm 90$ MeV \cite{BaBar:2007ceh}. This description aligns well with the $1 ^{3}D_{1}$ state of the pion in the current study in both formalisms.
While study \cite{Godfrey:1985xj} estimate the mass for this state to be nearly 100 MeV larger than the current work, studies \cite{Ishida:1986vn}, \cite{Ebert:2009ub} and \cite{Kojo:2022psi} estimate it near the mass of the present work. Ref. \cite{Ebert:2009ub} and \cite{Kojo:2022psi} also predict this state to be $\rho(1570)$ resonance. 

The $\rho(2150)$ has a similar $I^{G}(J^{PC})$ value as the $\rho(1570)$ resonance and has an average mass of 2150 MeV. It has been observed in various processes, including $e^{+}e^{-}$ annihilation, $\bar{p}p\rightarrow\pi\pi$, S-channel $\bar{N}N$ and $\pi^{-}p\rightarrow\omega\pi^{0}n$ \cite{ParticleDataGroup:2022pth}. This resonance fits very well with the $2 ^{3}D_{1}$ state of the pion in the current study in both formalisms. Ref. \cite{Godfrey:1985xj} and \cite{Ishida:1986vn} estimate the mass of this state to be a few MeV higher, while ref. \cite{Vijande:2004he}, \cite{Ebert:2009ub} and \cite{Kojo:2022psi} estimate it to be at least 100 MeV lower. 

\subsubsection*{\textbf{I. $^{3}D_{3}$ State}}

The $\rho_{3}(1690)$ state is a well-defined pionic resonance with an average mass of $1688.8\pm2.1$ MeV and a $I^{G}(J^{PC}) = 1^{+}(3^{--})$. The mass of this resonance for $2\pi$ mode, $K\bar{K}$ and $K\bar{K}\pi$ mode, $(4\pi)^{+-}$ mode, $\omega\pi$ mode and $\eta\pi^{+}\pi^{-}$ mode has a value $1686\pm4$ MeV, $1696\pm4$ MeV, $1686\pm5$ MeV, $1681\pm7$ MeV and $1682\pm12$ MeV respectively \cite{ParticleDataGroup:2022pth}. Similarly, the decay widths of these modes are $186\pm14$ MeV, $204\pm18$ MeV, $129\pm10$ MeV, $190\pm40$ MeV and $126\pm40$ MeV. This mass description matches the $1 ^{3}D_{3}$ state of the present work in semi-relativistic formalism very well. The compared studies also predict this state to be $\rho_{3}(1690)$ resonance. 

The $2 ^{3}D_{3}$ state of the present work has two resonances as potential candidates, namely $X(2110)$ and $\rho_{3}(2250)$. The $X(2110)$ resonance has been observed in the $10,16\pi^{-}p\rightarrow\bar{p}pn$ process with an average mass of $2110\pm10$ MeV and a decay width of $330\pm20$ MeV \cite{Bari-Bonn-CERN-Daresbury-Glasgow-Liverpool-Milan-Vienna:1978upm}. However, this resonance has a $I^{G}(J^{PC})$ value, which is not very well defined at $1^{+}(3^{-?})$. On the other hand, the $\rho_{3}(2250)$ has a very well defined $I^{G}(J^{PC})$ value, $1^{+}(3^{--})$. This resonance has been observed in various processes but no average mass value is decided by PDG \cite{ParticleDataGroup:2022pth}. This resonance has a mass ranging from 2090 MeV to 2300 MeV. Hence, $X(2110)$  resonance can be favored more as a potential candidate. Except ref. \cite{Godfrey:1985xj} and \cite{Ishida:1986vn}, all the compared studies estimate the mass of the $2 ^{3}D_{3}$ state to be lower by at least 100 MeV than the present work. Ref. \cite{Ishida:1986vn} predicts this state to be $\rho_{3}(2250)$ resonance, while ref. \cite{Vijande:2004he} predicts it to be $\rho_{3}(1990)$. The D-wave pion also follows the same trend as the S-wave and P-wave pions in the matter of mass of the radially excited state. 

\subsubsection*{\textbf{J. $^{3}F_{2}$ State}}

The $X(1870)$ resonance was observed in the production of $G(1590)$ and other mesons decaying into $\eta$ pairs by $100GeV/c\pi^{-}$ on protons, specifically in the $100\pi^{-}p\rightarrow 2\eta X$ process \cite{Serpukhov-Brussels-LosAlamos-AnnecyLAPP:1985qjh}. This resonance is not well established and has a $I^{G}(J^{PC})$ value that is not very well defined, $?^{?}(2^{??})$. This resonance has an average mass of $1870\pm40$ MeV and a decay width of $250\pm30$ MeV. The mass and J value of this resonance match the $1 ^{3}F_{2}$ pion state of the present work. While ref. \cite{Ebert:2009ub} estimates a lower mass for the $1 ^{3}F_{2}$ state, ref. \cite{Godfrey:1985xj} and \cite{Ishida:1986vn} estimate a higher mass for the state.  

\subsubsection*{\textbf{K. $^{3}F_{3}$ State}}

The $a_3(1875)$ resonance has been detected in the process $18.3\pi^{-}p \rightarrow \pi^{+}\pi^{-}\pi^{-}p$, with an average mass of $1874 \pm 43 \pm 96$ MeV and an average decay width of $385 \pm 121 \pm 114$ MeV \cite{Chung:2002pu}. This resonance is characterized by a well-defined quantum number $I^{G}(J^{PC})$ value of $1^{-}(3^{++})$, making it a strong candidate for the $1 ^{3}F_{3}$ state of the pion in the current study. All the compared studies estimate a higher mass for this state. Ref. \cite{Ishida:1986vn} predicts this state to be the $a_{3}(2050)$ resonance and ref. \cite{Ebert:2009ub} predicts this state to be the $a_{3}(1875)$ resonance.

The $a_{3}(2275)$ resonance has been observed in the partial wave analysis of $\bar{p}p\rightarrow\eta\eta\pi^{0}$, especially in the $1.96-2.41 \bar{p}p$ process with an average mass of $2275\pm35$ MeV and an average decay width of $350^{+100}_{-50}$ MeV \cite{Anisovich:2001pp}. This resonance can be predicted for the $2 ^{3}F_{3}$ state of the pion in the semi-relativistic formalism of the current study. Only ref. \cite{Ebert:2009ub} estimates the mass for $2 ^{3}F_{3}$, which is 150 MeV less than the present study and it predicts it to be the $a_{3}(2070)$.

\subsubsection*{\textbf{L. $^{3}F_{4}$ State}}

The $1 ^{3}F_{4}$ pion state in the present work has two resonances as potential candidates: $a_{4}(1970)$ and $X(2000)$. The $a_{4}(1970)$ resonance, which was earlier known as $a_{4}(2040)$ resonance, has an average mass of $1967\pm16$ MeV and an $I^{G}(J^{PC})$ value of $1^{-}(4^{++})$. This resonance was observed in $190\pi^{-}p\rightarrow\pi^{-}\pi^{+}\pi^{-}p$ \cite{COMPASS:2018uzl}, $191\pi^{-}p\rightarrow\eta^{'}\pi^{-}p$ \cite{COMPASS:2014vkj}, $18\pi^{-}p\rightarrow\omega\pi^{-}\pi^{0}p$ \cite{E852:2004rfa} and many such processes with an average decay width of $324^{+15}_{-18}$ MeV. At the same time, the $X(2000)$ resonance does not have a well-defined $I^{G}(J^{PC})$ value, $?^{?}(4^{++})$. This resonance has been observed in $\pi^{-}p\rightarrow K_{S}^{0}K_{S}^{0}MM$ process with an average mass of $1998\pm3\pm5$ MeV and an average decay width that is less than 15 MeV. These characteristics make both the resonances potential candidates for the $1 ^{3}F_{4}$ pion state in the present work under semi-relativistic formalism. The compared studies also predict similar mass values for this state and predict $a_{4}(1970)$ for this state. 

The $1 ^{3}F_{4}$ pion state in the present work under the non-relativistic formalism shows good proximity with the $X(2360)$ resonance. This resonance has been observed in the partial wave analysis of the system produced at low four momentum transfer in the $\pi^{-}p\rightarrow p\bar{p}n$ reaction at 18 GeV with a mass of $2360\pm10$ MeV and a decay width of $430\pm30$ MeV \cite{Rozanska:1979ub}. The $I^{G}$ value of this resonance is not yet defined but the $J^{P}$ value is $4^{+}$. Hence, this resonance can be seen as a potential candidate for the $1 ^{3}F_{4}$ pion state. Only ref. \cite{Ebert:2009ub} has estimated the mass of this state and gives a lower mass value than the present work.

\subsubsection*{\textbf{M. $^{1}G_{4}$ State}}

The $\pi_{4}(2250)$ resonance has all the characteristics to be a potential candidate for the $1 ^{1}G_{4}$ pion state in the present work. This resonance has been observed with mass $2250\pm15$ and decay rate $215\pm25$ in the $2.0\bar{p}p\rightarrow3\pi^{0},\pi^{0}\eta,\pi^{0}\eta^{'}$ process \cite{Anisovich:2001pn}. The $I^{G}(J^{PC})$ values of this resonance match those of the $1 ^{1}G_{4}$ pion state. Ref. \cite{Ebert:2009ub} estimates a lower mass for this state but does not predict any state for it.

\subsubsection*{\textbf{N. $^{3}G_{3}$ State}}

The $1 ^{3}G_{3}$ pion state also has two resonances as potential candiates: $\rho_{3}(2250)$ and $X(2110)$. The former resonance has been observed in $\bar{p}p\rightarrow\pi\pi,K\bar{K}$, S-Channel $\bar{N}N$ other such processes with $I^{G}(J^{PC})$ values $1^{+}(3^{--})$ \cite{ParticleDataGroup:2022pth}. PDG does not estimate a central mass value for this resonance but this resonance has been observed with masses ranging from 2150 MeV to 2248 MeV in the observed processes. The latter resonance, $X(2110)$, has a half-defined $I^{G}(J^{PC})$ value of $1^{+}(3^{-?})$ and has been observed with an average mass of $2110\pm10$ MeV and decay width of $330\pm20$ \cite{Bari-Bonn-CERN-Daresbury-Glasgow-Liverpool-Milan-Vienna:1978upm}. The $1 ^{3}G_{3}$ pion state in the present study does match the description for these two resonances. But since, no central mass value for $\rho_{3}(2250)$ is available, $X(2110)$ seems the more potent candidate. Ref. \cite{Godfrey:1985xj} estimates a higher mass for this state, while ref. \cite{Ebert:2009ub} estimates a lower mass and predicts this state to be the $\rho_{3}(1990)$ resonance.

\subsubsection*{\textbf{O. $^{3}G_{5}$ State}}

Similarly, the $1 ^{3}G_{5}$ pion state also has two resonances as potential candidates, $\rho_{5}(2350)$ and $X(2440)$. The former resonance has been observed in $\pi^{-}p\rightarrow\omega\pi^{0}n$, $\bar{p}p\rightarrow\pi\pi,K\bar{K}$, S-Channel $\bar{N}N$ and $\pi^{-}p\rightarrow K^{+}K^{-}n$ processes with $I^{G}(J^{PC})$ values $1^{+}(5^{--})$ \cite{ParticleDataGroup:2022pth}. PDG does estimate an average mass of $2330\pm50$ MeV and an average decay width of $400\pm100$ MeV. This resonance was earlier known as the $U_{1}(2400)$ resonance. The latter resonance, $X(2440)$ has a half-defined $(J^{PC})$ value, $(5^{-?})$ and an undefined $I^{G}$ value. It has been observed with a mass of $2440\pm10$ MeV and a decay width of $310\pm20$ MeV in the partial wave analysis of the system produced at low four momentum transfer in the $\pi^{-}p\rightarrow p\bar{p}n$ reaction at 18 GeV \cite{Rozanska:1979ub}. The $1 ^{3}G_{5}$ pion state for semi-relativistic formalism in the present study does match the description for these two resonances very well. Ref. \cite{Godfrey:1985xj} and \cite{Ebert:2009ub} predict slightly lower masses for this state and ref. \cite{Ebert:2009ub} also predicts this state to be $\rho_{5}(2300)$, which is now known as $\rho_{5}(2350)$.

\subsection{Tetraquark}

Since no PDG data for light strange tetraquarks is available, the two-meson threshold limit is used as a reference for comparison. Since the present study cannot distinguish between u and d quarks, cases for isospin, I = 0, 1 and 2, are considered. 

\subsubsection*{\textbf{A. $^{1}S_{0}$ State}}

The $1 ^{1}S_{0}$ state has three resonances as potential candidates under consideration: $f_{0}(980)$, $a_{0}(980)$, $a_{0}(980)$ and $X(1070)$. The $f_{0}(980)$ resonance has a mass of $990\pm20$ MeV and has been observed in numerous processes with decay widths ranging between 10 and 100 MeV. Similarly, the $a_{0}(980)$ resonance has an average mass of $980\pm20$ MeV with decay widths ranging between 50 and 100 MeV \cite{ParticleDataGroup:2022pth}. The $X(1070)$ resonance was observed in the $40\pi^{-}p\rightarrow K^{0}_{S}K^{0}_{S}n+m\pi^{0}$ process with an average mass of $1072\pm1$ MeV and a decay width of $3.5\pm0.5$ MeV \cite{Vladimirsky:2008zz}. While all three resonances have a $J^{PC}$ value of $0^{++}$, the $I^{G}$ value is $0^{+}$ for $f_{0}(980)$, $1^{-}$ for $a_{0}(980)$, and undefined for $X(1070)$. Ref. \cite{Vijande:2009ac} predicts this state to be the $f_{0}(980)$ resonance with a mass of 999 MeV. Ref. \cite{Santopinto:2006my} identifies it as the $f_{0}(600)$ resonance with a mass of 550 MeV. Ref. \cite{Ebert:2008id} predicts it to be the $f_{0}(600)$ and $f_{0}(1370)$ resonances, with masses of 596 MeV and 1179 MeV, respectively. Ref. \cite{Agaev:2018fvz} predicts this state to be the $a_{0}(980)$ resonance with mass $991^{+29}_{-27}$. 
Ref. \cite{Zhao:2021jss}, on the other hand, estimates this state with a higher mass of 1431 MeV, 1812 MeV, 1676 MeV and 1812 MeV and predicts it to be the $f_{0}(1500)$, $f_{0}(1710)$ and $f_{0}(2020)$ resonance. Ref. \cite{Wang:2019nln} predicts the mass of this state to be $1860\pm110$ MeV. The present study estimates the decay width for the radial ground state of $^{1}S_{0}$ to be nearly 255 MeV in semi-relativistic formalism and 280 MeV for non-relativistic formalism, which is a bit higher than that of the resonances predicted here. While the major branching channel occurs to be $\rho+\text{3 gluons}$, this still leaves about 100 to 150 MeV of decay width unknown. The next major decay channel following that is $\pi+\text{2 photons}$. 

The $X(2540)$ is a very good candidate for the $2 ^{1}S_{0}$ tetraquark state in the ${\textbf{6}}-\bar{\textbf{6}}$ color configuration. This resonance has been observed in the $\gamma\gamma\rightarrow K_{S}^{0}K_{S}^{0}$ process with a mass of $2539\pm14^{+38}_{-14}$ and a decay width of $274^{+77+126}_{-61-123}$ MeV \cite{Belle:2013eck}. The $I^{G}(J^{PC})$ value of this resonance is $0^{+}(0^{++})$ and matches the state very well. 

The $2 ^{1}S_{0}$ state has very similar characteristics to the $f_{0}(2330)$ resonance. This resonance has been observed with mass ranging from 2312 MeV to 2419 MeV and decay width ranging from 65 MeV to 274 MeV in $J/\psi(1S)\rightarrow\gamma\eta\eta^{'}$ \cite{BESIII:2022iwi},  $J/\psi\rightarrow\gamma\eta^{'}\eta^{'}\rightarrow4\gamma2(\pi^{+}\pi^{-}) $ \cite{BESIII:2022zel}, $J/\psi(1S)\rightarrow\gamma(\pi\pi,K\bar{K}) $ \cite{Rodas:2021tyb}, $J/\psi(1S)\rightarrow\gamma(\pi\pi,K\bar{K},\eta\eta,\omega,\phi) $ \cite{Sarantsev:2021ein}, $2.0\bar{p}p\rightarrow\pi\pi,\eta\eta$ \cite{Bugg:2004rj} and $\bar{p}p\rightarrow\pi\pi$ process \cite{Anisovich:2000ut}. For this state, ref. \cite{Zhao:2021jss} estimates mass at 1886 MeV, 1986 MeV, 2141 MeV and 2252 MeV and assigns resonances $f_{0}(2020)$, $f_{0}(2100)$ and $f_{0}(2200)$ to them. 

\subsubsection*{\textbf{B. $^{5}S_{2}$ State}}

For $1 ^{5}S_{2}$ state four resonance,$a_{2}(1990)$ $f_{2}(2000)$, $f_{2}(2010)$ and $a_{2}(2030)$ show good proximity. The resonances $a_{2}(1990)$ and $a_{2}(2030)$ have been discussed earlier. The $f_{2}(2000)$ resonance has been detected in a study \cite{Anisovich:2000ut} with a mass $2001\pm10$ MeV and a decay width of $312\pm32$ MeV. This resonance has also been detected in the $\bar{p}p\rightarrow\pi\pi$ process with a mass of 1996 Mev and a decay width of 134 MeV \cite{Hasan:1994he}. The $f_{2}(2010)$ resonance has been observed in the $22\pi^{-}p\rightarrow\phi\phi n$ process with an average mass of $2011^{+60}_{-80}$ MeV and a decay width of $202\pm60$ MeV \cite{Etkin:1987rj}. This resonance is a well-established particle and has the $I^{G}(J^{PC})$ value of $0^{+}(2^{++})$. Ref. \cite{Zhao:2021jss} estimates the mass of $1 ^{5}S_{2}$ to be 1936 MeV and 1978 Mev and predicts $X_{2}(1930)$ and $X_{2}(1980)$ resonance for it. Ref. \cite{Ebert:2008id} predicts the mass of $1 ^{5}S_{2}$ state to be 1915 MeV and assigns $f_{2}(1910)$ and $f_{2}(1950)$ to it. Ref. \cite{Wang:2019nln} predicts the mass of this state to be $2130\pm100$ MeV. Ref. \cite{Zhao:2021jss} estimates the mass of $2 ^{5}S_{2}$ less than the present work and predicts $f_{2}(2340)$ and $f_{2}(2300)$ resonance for it.

\subsubsection*{\textbf{C. $^{1}P_{1}$ State}}

The $\omega(2290)$ resonance has been observed in the partial wave analysis of $\bar{p}p\rightarrow\bar{\Lambda}\Lambda$ with an average mass of $2290\pm20$ MeV and decay width $275\pm35$ MeV \cite{Bugg:2004rj}. Similarly, the $\omega(2330)$ resonance has been observed in the photon diffractive dissociation to $\rho\rho\pi$ and $\rho\pi\pi\pi$ states in $25-50\gamma p\rightarrow\rho^{\pm}\rho^{0}\pi^{\mp}$ with an average mass of $2330\pm30$ and decay width of $435\pm75$ MeV \cite{OmegaPhoton:1988guj}. Both of these resonances have the $I^{G}(J^{PC})$ value $0^{-}(1^{--})$ and can be considered potential candidates for the $1 ^{1}P_{1}$ tetraquark state in both color configurations. Ref. \cite{Xin:2022qnv} estimates the mass of the P-wave vector state for all light tetraquark between $1980\pm110$ MeV and $2890\pm100$ MeV, but does not predict any state or resonance for it. 

\section*{Conclusion}

In conclusion, the mass spectra of all light tetraquark have been estimated in non-relativistic and semi-relativistic (correction to kinetic energy) formalism. To do so, the mass spectra of pions are calculated and the obtained parameters are used in the in the calculation of diquark and tetraquark masses. The computed states have corresponding $J^{PC}$ values assigned to them. In order to study the decay properties of the all-light tetraquark, the annihilation model \cite{Kher:2018wtv}, spectator model \cite{Becchi:2020mjz,Becchi:2020uvq}, and rearrangement model \cite{Ali:2019roi} were employed. For various spins and parity states, regge plots for pions and all light tetraquarks have been plotted. Given the lack of reference material, two-meson thresholds in both formalisms for tetraquarks are also estimated. And finally, more than 30 resonances have been investigated as potential candidates for S, P, D, F and G wave pion states. Similarly, 11 resonances have been investigated as potential candidates for S and P wave tetraquark states. {Several resonances, such as $\rho(1570)$, $a_{0}(980)$, and $f_{0}(980)$, have been proposed as potential candidates for multiple states due to their mass descriptions aligning with more than one possible state. To achieve a clearer distinction between these candidates, a more in-depth investigation into properties like decay width, magnetic moment, and other characteristics is necessary. These aspects will be thoroughly examined in future studies to better understand and differentiate these resonances. }

The present investigation provides the foundation for future studies employing the formalism described here to investigate tetraquarks with various quark flavors and understand the physics of unflavored mesons and tetraquark in the 0 to 3 GeV mass range. These findings could potentially serve as valuable input and comparative data for upcoming facilities such as PANDA \cite{PANDA:2016scz,PANDA:2016fbp,Singh:2016hoh,Singh:2019jug,PANDA:2018zjt,PANDA:2023ljx,PANDA:2021ozp} and others which focus on in-depth analyses of resonances involving light quarks.

\section{Data Availability Statement} The datasets generated during and/or analysed during the current study are available from the corresponding author on reasonable request.

%
%


\end{document}